\documentclass[12pt]{article}
\usepackage{graphicx}
\usepackage{amssymb,epsfig}
\usepackage{lscape,graphics,amsmath}
\usepackage{latexsym,amsfonts}
\usepackage{axodraw}
\usepackage{cite}
\usepackage[english]{babel}

\newcounter{diag1}
\newcounter{diag3}
\newcounter{diag4}
\begin{document}

\begin{center}
{\Large \bf Coherence length of neutrino oscillations in quantum field-theoretical  approach}\\
\vspace{4mm} Vadim O. Egorov$^{1,2}$, Igor P.~Volobuev$^1$\\
\vspace{4mm} $^1$Skobeltsyn Institute of Nuclear Physics, Moscow State
University,
\\ 119991 Moscow, Russia\\
$^2$Faculty of Physics, Moscow State University, 119991 Moscow, Russia
\end{center}

\vspace{0.5cm}
\begin{abstract}
We consider a novel quantum field-theoretical approach to the
description  of processes passing at finite space-time intervals
based on the Feynman diagram technique in the coordinate
representation. The most known processes of this type are neutrino
and neutral kaon oscillations. The experimental setting of these
processes requires one to adjust the rules of passing to the
momentum representation in the Feynman diagram technique in
accordance with it, which leads to a modification of the Feynman
propagator in the momentum representation. The approach does not
make use of wave packets, both initial and final particle states
are described by plane waves, which simplifies the calculations
considerably. We consider neutrino oscillation processes, where
the neutrinos are produced in three-particle weak decays of nuclei
and detected in the charged-current interaction with nuclei or in
the charged- and neutral-current interactions with electrons.
Particular  examples are considered  and it is shown that the
momentum spread of the produced neutrinos and the energy
dependence of the differential cross section of the detection
process result in the suppression of neutrino oscillation, which
is characterized by a coherence length specific for a pair of
production and detection processes. This coherence length turns
out to be much less than the coherence length in the standard
quantum-mechanical approach defined by the  quantum uncertainty of
neutrino momentum.
\end{abstract}

\section{Introduction}
The Standard Model allows one to describe a great amount of
different elementary particle interaction processes with a high
accuracy in the framework of the perturbative S-matrix formalism
and the  Feynman diagram technique. However, there is a number of
phenomena which cannot be described in the framework of the
standard perturbation theory. In particular, these are strange
neutral meson oscillations and neutrino oscillations, which take
place at finite macroscopic space and time intervals. These
phenomena are described either in the quantum mechanical approach
in terms of plane waves
\cite{Pais:1955sm,Pontecorvo:1957cp,Gribov:1968kq,Belusevic:1998pw,Giunti:2007ry,Bilenky:2010zza,Petcov}
or in the QM or QFT approaches in terms of wave packets
\cite{Giunti:1993se,Grimus:1996av,Beuthe:2001rc,Cohen:2008qb,Lobanov:2015esa}.
The first one is based on the notion of the states with definite
flavour (definite strangeness) which are superpositions of the
states with definite mass. It is postulated that it is the flavour
states that are produced in the weak interaction, and their
evolution in time underlies the oscillations. However, in the
plane wave approximation, the production of states without
definite mass leads to  violation of energy-momentum conservation,
which was widely discussed in the literature
\cite{Giunti:1993se,Grimus:1996av,Beuthe:2001rc,Cohen:2008qb,Lobanov:2015esa}.
This problem can be solved in the framework of the wave-packet
treatment \cite{Giunti:2007ry}, but the price  is an essential
complication of the corresponding calculations.

An alternative quantum field-theoretical description put
forward in \cite{Giunti:1993se}  and developed in
\cite{Grimus:1996av,Beuthe:2001rc} explains the neutrino
oscillations by  interference of the amplitudes of processes
mediated by  different virtual neutrinos with definite masses. In
the framework of this description there are no problems with
energy-momentum conservation, but the calculations of amplitudes
turn out to be rather complicated because of the necessity to use
wave packets in order to describe a localization of particles or
nuclei. The calculation procedure is essentially different from
the standard calculations in the Feynman diagram technique in the
momentum representation. This is due to the standard S-matrix
formalism of QFT not being convenient for describing
processes at finite distances and finite time intervals.

In what follows, we show that neutrino oscillations may be
consistently  described in the framework of quantum field theory
using only plane waves, which simplifies the calculations
considerably. Nevertheless, in the developed approach
energy-momentum is conserved as well. The idea of the novel
approach is to adjust the standard S-matrix formalism for
describing  the processes of finite duration. We consider the
processes of production and detection as a whole, use the Feynman
diagram technique in the coordinate representation to write down
the amplitude and then pass to the momentum representation in a
way, which corresponds to the experimental setting. Effectively it
leads to a modification of the Feynman propagator in the momentum
representation, while all the other Feynman rules in the momentum
representation are kept intact. The approach is based on two
papers by R. Feynman \cite{Feynman:1949hz,Feynman:1949zx} and
developed in  papers
\cite{Volobuev:2017izt,Egorov:2017qgk,Egorov:2017vdp}. In the
present paper, in the framework of the proposed approach, we
consider the processes of neutrino oscillations, where the
neutrinos produced in weak decays of nuclei are detected either in
the weak charged-current interaction with  nuclei or in both the
charged- and neutral-current interactions with  electrons.

 An important characteristic of neutrino oscillation processes
is the coherence length, which is the measure of fading out of the
oscillation pattern with distance. It appears in the
quantum-mechanical description of neutrino oscillation in terms of
wave packets due to the momentum uncertainty of the neutrino
states.  Meanwhile, in the framework of this approach one
considers only the neutrino states with the same expectation value
of momentum, which enters the expressions for the oscillation
lengths. This means that a beam of such neutrinos can be viewed as
a monochromatic one at the distances from the source much less
than the coherence lengths. However, the neutrinos produced in
three-particle weak decays of nuclei are not monochromatic, and
the spread of neutrino momenta can also affect the oscillation
pattern.

In the quantum field-theoretical approach to neutrino oscillations
under consideration there is no momentum uncertainty of neutrino
states, because all the particles, just like in the standard
Feynman diagram technique, are described by plane waves. For this
reason fading out of the oscillation pattern in this approach can
result only from the momentum spread of  the produced neutrinos.
In what follows we examine specific examples and show how the
coherence length appears in our approach and what are the
differences between our approach and the standard one.

\section{Neutrino oscillations in experiments with detection in the charged-current interaction only}

\subsection{Theory}

We work in the framework of the minimal extension of the Standard
Model  by the right neutrino singlets. The charged-current
interaction Lagrangian of the leptons takes the form
\begin{equation}\label{L_cc}
L_{\rm cc} = - \frac{g }{2\sqrt{2}}\left(\sum_{i,k = 1}^3 \bar l_i
\gamma^\mu (1 - \gamma^5)U_{ik}\nu_k   W^{-}_\mu + h.c.\right),
\end{equation}
where $l_i$ is the field of the charged lepton of the $i$-th
generation,  $U_{ik}$ denotes the PMNS-matrix, and  $\nu_k$ stands
for the field of the neutrino state with definite mass.

Let us consider a process, where a neutrino is emitted  and
detected in the charged-current interaction with nuclei. In
the lowest order of  perturbation theory the process is described
by the following diagram:
\vspace{0.3cm}
\begin{figure}[h]
\begin{center}
\ \ \ \includegraphics[width=0.3965\linewidth]{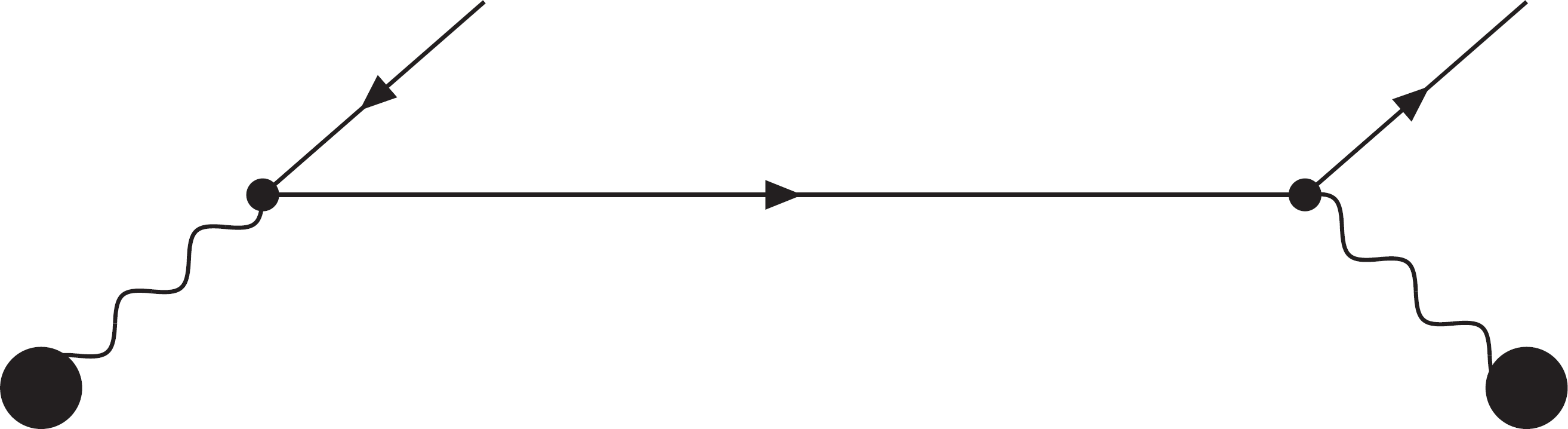}
\end{center}
\end{figure}
\vspace*{-2.961cm}
\begin{center}
\begin{picture}(197,87)(0,0)
\Text(70.0,94.0)[l]{$e^+ ( q )$}\ArrowLine(67.5,88.0)(40.5,64.5)
\Text(33.5,65.5)[r]{$x$} \Photon(13.5,41.0)(40.5,64.5){2}{3.0}
\Text(53.5,48.5)[r]{$W^+$} \Vertex (13.5,41.0){5} \Vertex
(40.5,64.5){2} \ArrowLine(40.5,64.5)(167.5,64.5) \Vertex
(167.5,64.5){2} \Text(104.8,70.5)[b]{$\nu_i ( p_{\rm n} )$}
\ArrowLine(167.5,64.5)(194.5,88.0) \Text(197.5,94.0)[l]{$e^- ( k
)$} \Text(175.0,64.5)[l]{$y$}
\Photon(167.5,64.5)(194.5,41.0){2}{3.0} \Vertex (194.5,41.0){5}
\Text(177.0,48.5)[r]{$W^+$}
\Text(330.0,60.5)[b]{\addtocounter{equation}{1}(\arabic{equation})}
\setcounter{diag1}{\value{equation}}
\label{diag1}
\end{picture}
\end{center}
\vspace{-1.349cm}
The points of production $x$ and detection $y$
are  supposed to be separated by a fixed macroscopic interval. The
intermediate neutrino mass eigenstate is a virtual particle and is
described by the propagator in the coordinate representation. All
three virtual neutrino mass eigenstates contribute to the
amplitude of the process, thus the amplitude of the process
corresponding to the diagram must be summed over all the
three neutrino mass  eigenstates, $i = 1,2,3$.

As it is customary in the Feynman diagram technique, we suppose
that the initial and final  nuclei and particles are
described by plane waves, i.e.\ they have definite momenta. Hence,
all the three virtual neutrino eigenstates have definite momenta
as well. Let us assign the 4-momenta of the particles as it is
shown in the diagram, namely, $q$ is the positron 4-momentum, $k$
is the electron 4-momentum and $p_{\rm n}$ is the intermediate virtual
neutrino 4-momentum. To be specific, we will suppose that the
virtual $W$-bosons are produced and absorbed in the interactions
with nuclei as follows: a nucleus ${^{A_1} _{Z_1} {\rm X}}$, which
will be referred to as  nucleus $1$, emits $W^+$-boson and turns
into the nucleus ${^{A_1} _{Z_1-1} {\rm X}}$, which will be referred to
as  nucleus $1^\prime$, and a nucleus ${^{A_2} _{Z_2} {\rm X}}$, which
will be referred to as  nucleus $2$, absorbs $W^+$-boson and turns
into the nucleus ${^{A_2} _{Z_2+1} {\rm X}}$, which will be referred to
as  nucleus $2^\prime$. Thereby the filled circles in the diagram
represent the matrix elements of the weak charged hadron current
$$j_\mu^{(1)} \left( {P^{(1)}, P^{(1^\prime)} } \right) =
 \left <^{A_1}_{Z_1 - 1} {\rm X} \right| j_\mu^{({\rm h})} \left| ^{A_1}_{Z_1} {\rm X} \right>,
 \quad j_\rho^{(2)} \left( {P^{(2)}, P^{(2^\prime)} } \right) =
 \left <^{A_2}_{Z_2 + 1} {\rm X} \right| j_\rho^{({\rm h})} \left| ^{A_2}_{Z_2} {\rm X} \right>,$$
corresponding to  nuclei $1,1^\prime$ and $2,2^\prime$; the nuclei
4-momenta are denoted by $P^{(l)} = (E^{(l)}, \vec P^{(l)})$,
$P^{(l^\prime)} = (E^{(l^\prime)}, \vec P^{(l^\prime)})$, $l=
1,2$.

The amplitude in the coordinate representation corresponding
to diagram (\arabic{diag1}) can be written out  using the Feynman
rules in the coordinate representation formulated, for example, in
textbook \cite{BOSH}. In order to pass to the momentum
representation one would have to integrate the amplitude with
respect to $x$ and $y$ over  Minkowski space, which would give the
corresponding matrix element of the S-matrix.

However, such an integration  would result in losing the
information about the space-time interval between the production
event and the detection event, because the experimental situation
in neutrino oscillation  experiments implies that the distance
between the production point and the detection point along the
neutrino propagation direction remains fixed. To generalize the
standard S-matrix formalism to the case of processes passing at
fixed distances, we have to modify the integration in such a way
that it would take into account a fixed distance between the
neutrino production  and detection points. This can be done by
introducing a delta function into the integral, which would fix
the distance between these points. However, in paper
\cite{Egorov:2017qgk} it was argued that it was more convenient to
fix the time interval between the production and detection events
by introducing the delta function $\delta(y^0 - x^0 - T)$ into the
integral, because, for a beam of neutrinos with the same momentum,
this is equivalent to fixing the distance between these events in
accordance with the formula $T = L p^0/|\vec p|$, which is often
used in describing neutrino oscillation processes
\cite{Giunti:2007ry}.

Having fixed the time interval between the events of production and
detection,  we integrate the amplitude with respect to $x$
and $y$ over Minkowski space. Thus, just like in the standard
S-matrix formalism, we consider the process taking place
throughout Minkowski space-time, but the time interval between the
production and detection events is now fixed by the delta
function. This is equivalent to replacing the standard Feynman
fermion propagator in the coordinate representation $S^{\rm c}_i(y-x)$
by $S^{\rm c}_i(y-x)\delta(y^0 - x^0 - T)$.

The Fourier transform of this expression  gives us the so-called
time-dependent propagator of the neutrino mass eigenstate $\nu_i$
in the momentum representation, defined by the relation:
\begin{equation}\label{prop_T_mom}
S^{\rm c}_i(p,T) = \int dx\, e^{ip x} S^{\rm c}_i(x)\,  \delta(x^0 - T).
\end{equation}

This integral can be evaluated exactly
\cite{Volobuev:2017izt,Egorov:2017qgk}:
\begin{equation}
\label{spin_prop}S^{\rm c}_i \left( {p,T} \right) = i \, \frac{{\hat p -
\gamma _0 \left( {p^0  - \sqrt {\left( {p^0 } \right)^2 + m_i^2 -
p^2} } \right) + m_i}}{{2\sqrt {\left( {p^0 } \right)^2 + m_i^2 -
p^2} }} \, e^{i\left( {p^0  - \sqrt {\left( {p^0 } \right)^2 +
m_i^2 - p^2} } \right)T}\,,
\end{equation}
where $m_i$ is the mass of $i$-th neutrino mass eigenstate and the
standard notation $\hat p = \gamma_\mu p^\mu $ is used. The
inverse Fourier transformation of this time-dependent propagator
is well defined, which allows us to retain the standard Feynman
diagram technique in the momentum representation just by replacing
the Feynman propagator by the time-dependent propagator.

In paper \cite{Grimus:1996av} it was rigorously proved  that virtual particles
propagating at large macroscopic distances (or, equivalently,
propagating over macroscopic times) are almost on the mass shell,
which means that $|p^2 - m_i^2|/ \left( {p^0 } \right)^2 \ll 1$.
This is in accord with the structure of time-dependent propagator
(\ref{spin_prop}).  As it was discussed in \cite{Egorov:2017vdp},
formally the amplitude of a process with such a propagator
corresponds to the instant registration. The process itself is
considered to take time $T$ exactly. It reality, however, the
registration has a non-zero duration $\Delta t$, and the amplitude
constructed with
 time-dependent propagator (\ref{spin_prop}) should be
interpreted as the amplitude per unit time. In order to find the
amplitude of a realistic process with the detection time $\Delta
t$ one must integrate the time-dependent amplitude with respect to
$T$ from $T - {{\Delta t} \mathord{\left/ {\vphantom {{\Delta t}
2}} \right. \kern-\nulldelimiterspace} 2}$ to $T + {{\Delta t}
\mathord{\left/ {\vphantom {{\Delta t} 2}} \right.
\kern-\nulldelimiterspace} 2}$. It reduces to the integration of
  propagator (\ref{spin_prop}) only, which gives
\begin{equation}
\int\limits_{T - \Delta t/2}^{T + \Delta t/2} {S_i^{\rm c} \left( {p ,t}
\right)dt}  = S_i^{\rm c} \left( {p ,T} \right) \frac{{\sin \alpha
}}{\alpha } \Delta t, \quad \alpha = \left( {p^0  - \sqrt {\left(
{p^0 } \right)^2 + m_i^2 - p^2} } \right)\frac{\Delta t}{2}.
\end{equation}
For large $\alpha \gg 1$ this integral is close to zero, and we
can expect that the amplitude will be essentially non-zero only
for those particles, for which $\alpha \simeq 0$. In this case the
amplitude with the registration time ${\Delta t}$ is proportional
to ${\Delta t}$, and the amplitude with $T$ fixed can really be
viewed as the amplitude of the registration per unit time.

The registration time interval $\Delta t$ is macroscopically
large, which means that the factor $\left( {p^0  - \sqrt {\left(
{p^0 } \right)^2 + m_i^2 - p^2} } \right)$ should be very small.
The latter is the expression of the fact that the virtual neutrino
is almost on the mass shell. Thus,  our approach actually gives
another proof of the  Grimus-Stockinger theorem
\cite{Grimus:1996av}.  Applying this result to time-dependent
propagator  (\ref{spin_prop}), i.e  neglecting  $|p^2 - m_i^2|/
\left( {p^0 } \right)^2 \ll 1$ everywhere, except in the
exponential, where it is multiplied by the macroscopic time $T$,
we get
\begin{equation} \label{spin_prop_1}
S_i^{\rm c} \left( {p,T} \right) = i\frac{{\hat p + m_i}}{{2p^0 }}e^{ - i\frac{{m_i^2  - p^2}}{{2p^0 }}T}.
\end{equation}
It is this expression that will be used for the calculations
hereinafter.

Now we are in a position to write down the amplitude in the
momentum representation corresponding to  diagram (\arabic{diag1})
in the case, when the time difference $y^0 - x^0$ between the
events of production and detection is fixed and equal to $T$.
Since the momentum transfer in both the production and detection
processes is small, one can use the approximation of Fermi's
interaction. Using the time-dependent fermion field propagator
(\ref{spin_prop_1}), where we retain the neutrino masses only in the
exponential, we arrive at the amplitude summed over all the three
neutrino mass eigenstates:
\begin{eqnarray}\label{amp_osc}
M &=&  - i\frac{{G_{\rm F}^{\,2} }}{4{p_{\rm n}^0}}\sum\limits_{i = 1}^3 {\left| {U_{1i} } \right|^2 e^{-i\frac{m_i^2 - p_{\rm n}^2}{ 2 p_{\rm n}^0} T} } \times \\
& & \times j_\rho ^{(2)} \left( {P^{(2)}, P^{(2^\prime)} } \right) \bar u
\left( k \right)\gamma ^\rho \left( {1 - \gamma ^5 } \right) \hat p_{\rm n}\gamma ^\mu \left( {1 - \gamma ^5 } \right)v
\left( q \right)j_\mu ^{(1)} \left( {P^{(1)}, P^{(1^\prime)} } \right) . \nonumber
\end{eqnarray}
Here and below we omit the fermion polarization indices for
simplicity.

The squared modulus of the amplitude, averaged with respect to the
polarizations of the incoming nuclei and summed over the
polarizations of the outgoing particles and nuclei (the operation
of averaging and summation is denoted by the angle brackets),
factorizes in the approximation of massless neutrinos as follows:
\begin{eqnarray} \label{sqr_amp}
\left\langle {\left| M \right|^2 } \right\rangle  &=& \left\langle
{\left| M_1 \right|^2 } \right\rangle \left\langle {\left| M_2
\right|^2 } \right\rangle \frac{{1}}{{4 \left( { p^0_{\rm n}}
\right)^2 }} \left[ {1 - 4\sum\limits_{\scriptstyle i,k = 1 \hfill
\atop \scriptstyle i < k \hfill}^3 {\left| {U_{1i} } \right|^2
\left| {U_{1k} } \right|^2 \sin ^2 \left( {\frac{{m_i^2  - m_k^2
}}{{4 p^0_{\rm n}}}T} \right)} } \right] ,\\
\label{sqr_M1} \left\langle {\left| M_1 \right|^2 } \right\rangle
&=& 4G_{\rm F}^{\,2} \left( -{g^{\mu \nu } \left( {p_{\rm n}q} \right) +\left(
{p_{\rm n}^\mu  q^\nu   + q^\mu  p_{\rm n}^\nu  } \right) +
i\varepsilon ^{\mu \nu \alpha \beta } p_{{\rm n} \alpha}  q_\beta  } \right) W_{\mu \nu }^{(1)} , \\
\left\langle {\left| M_2 \right|^2 } \right\rangle
&=& 4G_{\rm F}^{\,2} \left(- {g^{\rho \sigma } \left( {p_{\rm n} k} \right) +
\left( {p_{\rm n}^\rho  k^\sigma   + k^\rho  p_{\rm n}^\sigma  } \right) -
i\varepsilon ^{\rho \sigma \alpha \beta } p_{{\rm n} \alpha}  k_\beta  }
\right) W_{\rho \sigma }^{(2)},
\end{eqnarray}
where the nuclear tensors $ W_{\mu \nu }^{(1)}$, $W_{\rho \sigma
}^{(2)}$  characterizing the interaction of  nuclei $1$ and $2$
with the virtual $W$-bosons are defined as
\begin{equation}\label{nucl_tensor}
 W_{\alpha \beta }^{(l)} = W_{\alpha \beta }^{(l,{\rm S})} + iW_{\alpha \beta }^{(l,{\rm A})} =
 \left\langle {j_\alpha ^{(l)} \left( {j_\beta ^{(l)} } \right)^ +  } \right\rangle , \quad l=1,2,
\end{equation}
their symmetrical parts $W_{\alpha \beta }^{(l,{\rm S})}$ being real
and the anti-symmetrical ones $iW_{\alpha \beta }^{(l,{\rm A})}$ being
imaginary.

Our next step is to find the differential probability of the
process,  where the intermediate neutrino momentum $p_{\rm n}$ is fixed
by the experimental setting. Let us denote the 4-momentum $p$:
$(p)^2 = 0$, the vector $\vec p$ satisfies the energy-momentum
conservation in the production vertex and is directed from the
source to the detector. According to the prescription formulated
in papers \cite{Volobuev:2017izt,Egorov:2017qgk,Egorov:2017vdp} we
multiply the squared modulus of the amplitude (\ref{sqr_amp}) by
the delta function of energy-momentum conservation $(2\pi )^4
\delta (P^{(1)} + P^{(2)} - P^{(1^\prime)} - P^{(2^\prime)} - q -
k)$, substitute $p$ instead of $p_{\rm n}$ everywhere in the amplitude
and multiply the result by the delta function $2\pi \delta (
P^{(1)} - P^{(1^\prime)} - q - p )$, which fixes the virtual
neutrino momentum, and integrate it with respect to the phase
volume of the final particles and nuclei. Besides this, now, when
the virtual neutrino momentum is fixed, one can pass from the time
interval $T$ to the distance travelled by the neutrino $L$
according to the formula $T = {{Lp^0 } \mathord{\left/
 {\vphantom {{Lp^0 } {\left| {\vec p} \right|}}} \right.
 \kern-\nulldelimiterspace} {\left| {\vec p} \right|}}$. Thus, we
 arrive at the differential probability, which also factorizes:
\begin{eqnarray}
\frac{{d^3 W}}{{d^3 p}} &=& \frac{1}{{2E^{(1)} 2E^{(2)} }}\int {\frac{{d^3 k}}
{{\left( {2\pi } \right)^3 2k^0 }}\frac{{d^3 q}}{{\left( {2\pi } \right)^3 2q^0 }}
\frac{{d^3 P^{(1')} }}{{\left( {2\pi } \right)^3 2E^{(1')} }}\frac{{d^3 P^{(2')} }}
{{\left( {2\pi } \right)^3 2E^{(2')} }}\left. {\left\langle {\left| M \right|^2 } \right\rangle } \right|_{\scriptstyle p_{\rm n}  = p \hfill \atop
  \scriptstyle T = {{Lp^0 } \mathord{\left/
 {\vphantom {{Lp^0 } {\left| {\vec p} \right|}}} \right.
 \kern-\nulldelimiterspace} {\left| {\vec p} \right|}} \hfill}
  \times } \nonumber \\
& & \times \left( {2\pi } \right)^4 \delta \left( {P^{(1)}  + P^{(2)}  - P^{(1')}  - P^{(2')}  - q - k} \right)
2\pi \delta \left( {P^{(1)}  - P^{(1')}  - q - p} \right) =  \nonumber \\
\label{dif_prob_osc} &=& \frac{{d^3 W_1 }}{{d^3 p}}W_2 P_{ee} \left( {L} \right).
\end{eqnarray}
Here
\begin{equation}\label{dif_W_1}
\frac{{d^3 W_1 }}{{d^3 p}} = \frac{1}{{2E^{(1)} }}
\frac{1}{{\left( {2\pi } \right)^3 2p^0 }} \int
{\frac{{d^3 q}} {{\left( {2\pi } \right)^3 2q^0 }}
\frac{{d^3 P^{(1')} }} {{\left( {2\pi } \right)^3 2E^{(1')}
}} \left. { \left\langle {\left| {M_1 } \right|^2 }
\right\rangle }\right|_{p_{\rm n}  = p} \left( {2\pi } \right)^4
\delta \left( {P^{(1)}  - P^{(1')}  - q - p} \right)}
\end{equation}
is the differential probability of decay of  nucleus $1$ into
nucleus $1^\prime$, a positron and a massless fermion with
momentum $\vec p$,
\begin{equation}
W_2  = \frac{1}{{2E^{(2)} 2p^0 }}\int {\frac{{d^3
k}}{{\left( {2\pi } \right)^3 2k^0 }} \frac{{d^3
P^{(2')} }}{{\left( {2\pi } \right)^3 2E^{(2')}}}
\left. {\left\langle {\left| {M_2 } \right|^2 } \right\rangle }\right|_{p_{\rm n}  = p} \left( {2\pi
} \right)^4
\delta \left( {P^{(2)}  + p - P^{(2')}  - k} \right)}
\end{equation}
is the probability of interaction of a massless fermion with
momentum $\vec p$ and  nucleus $2$ with the production of nucleus
$2^\prime$ and an electron, and we introduced a special notation
\begin{equation}\label{P_ee}
P_{ee} \left( {L} \right) = 1 - 4\sum\limits_{\scriptstyle i,k = 1 \hfill \atop
\scriptstyle i < k \hfill}^3 {\left| {U_{1i} } \right|^2 \left|
{U_{1k} } \right|^2 \sin ^2 \left( {\frac{{m_i^2  - m_k^2
}}{{4\left| {\vec p} \right|}}L} \right)}
\end{equation}
for the expression, which, in the standard approach, is called
the distance-dependent electron neutrino survival probability.
Thus, one finds that the differential probability of the whole
process is the product of the differential probability $\frac{{d^3
W_1 }}{{d^3 p }}$ of the production of a neutrino with a definite
momentum, the probability $W_2$ of its interaction in the detector
and the standard distance-dependent oscillating factor $P_{ee}
\left( {L} \right)$.

Finally we observe that the experimental situation fixes only  the
direction of the intermediate neutrino momentum, but not its
length. However, the considered process of the neutrino production
is a three-body decay, hence the neutrino momentum can have
different lengths in a given direction. In order to take into
account the neutrinos with all the possible momenta directed from
the source to the detector, one has to integrate the differential
probability (\ref{dif_prob_osc}) multiplied by $\left| {\vec p}
\right|^2$ with respect to $\left| {\vec p} \right|$ from $\left|
{\vec p} \right|_{\min }$ to $\left| {\vec p} \right|_{\max }$. In
what follows, we assume  nuclei $1$ and $2$ to be at rest and  put
their initial momenta $\vec P^{(1)}$, $\vec P^{(2)}$ equal to
zero. Then the lower limit of integration determined by the
threshold of the registration process and the upper one determined
by the energy-momentum conservation in the production vertex are
given by \cite{BK}:
\begin{equation}
\left| {\vec p} \right|_{\min }  = \frac{{\left( {M_{2'}  + m} \right)^2  - M_2^2 }}{{2M_2 }}, \quad
\left| {\vec p} \right|_{\max }  = \frac{{M_1^2  - \left( {M_{1'}  + m} \right)^2 }}{{2M_1 }}.
\end{equation}
Here $M_1$, $M_{1'}$, $M_2$, $M_{2'}$ are the masses of  nuclei
$1$, $1'$, $2$, $2'$, respectively, and $m$ is the electron mass.
As a result we arrive at the total probability of detecting an
electron in the process under consideration:
\begin{equation}\label{prob_osc_fin}
\frac{{dW}}{{d\Omega }} = \int\limits_{\left| {\vec p} \right|_{\min } }^{\left| {\vec p} \right|_{\max } }
{\frac{{d^3 W}}{{d^3 p}}\left| {\vec p} \right|^2 d\left| {\vec p} \right|}  =
\int\limits_{\left| {\vec p} \right|_{\min } }^{\left| {\vec p} \right|_{\max } }
{\frac{{d^3 W_1 }}{{d^3 p}}W_2 P_{ee} \left( {L} \right) \left| {\vec p} \right|^2 d\left| {\vec p} \right|} .
\end{equation}
In the next subsection we will apply this formula to specific
 neutrino oscillation processes.

\subsection{Specific examples}

Let us consider a few examples with two reaction of the solar
carbon cycle
$${^{15} {\rm O}} \to {^{15} {\rm N}} + e^ +   + \nu_i \quad {\rm and} \quad {^{13} {\rm N}} \to {^{13} {\rm C}} + e^ +   +
\nu_i\,.
$$
First, let us take the production process to be the decay of ${^{15} {\rm O}}$
and the detection to be performed by  chlorine-argon or
gallium-germanium detectors,
$$\nu_i + {^{37} {\rm Cl}} \to {^{37} {\rm Ar}} + e^- \quad {\rm and} \quad \nu_i + {^{71} {\rm Ga}} \to {^{71} {\rm Ge}} + e^- .$$
In nuclear physics, these reactions refer to the so-called
allowed transitions \cite{Bohr-Mottelson}. In this case one can
neglect the nucleon positions and momenta, and the nucleons decay
or interact as if they were at rest. Correspondingly, one can
neglect the dependence of the nuclear form-factors on the momentum
transfer \cite{Bohr-Mottelson}.  If we also neglect the possible
contribution of the excited states of the final nuclei, the
product of the differential probability of neutrino production and
the probability of neutrino detection can be approximated by the
function
\begin{eqnarray}\label{distr_1}
\frac{{d^3 W_1 }}{{d^3 p}}W_2  &=& C\sqrt {\left( {\left| {\vec p}
\right|_{\max }  - \left| {\vec p} \right|} \right)\left( {\left|
{\vec p} \right|_{\max }  - \left| {\vec p} \right| + 2m} \right)}
\left( {\left| {\vec p} \right|_{\max }  - \left| {\vec p} \right| + m} \right) \times \nonumber \\
& & \times \sqrt {\left( {\left| {\vec p} \right| - \left| {\vec p} \right|_{\min } } \right)\left( {\left| {\vec p} \right| -
\left| {\vec p} \right|_{\min }  + 2m} \right)} \left( {\left| {\vec p} \right| - \left| {\vec p} \right|_{\min }  + m} \right).
\end{eqnarray}
This approximation is rather rough.  Nevertheless, it is
sufficient to demonstrate that, in the approach under
consideration, the coherence length of neutrino oscillations
arises due to the neutrino momentum spread and is defined by the
spectral characteristics of the production and detection
processes.

\begin{figure}[htb]
\center{\includegraphics[width=0.5\linewidth]{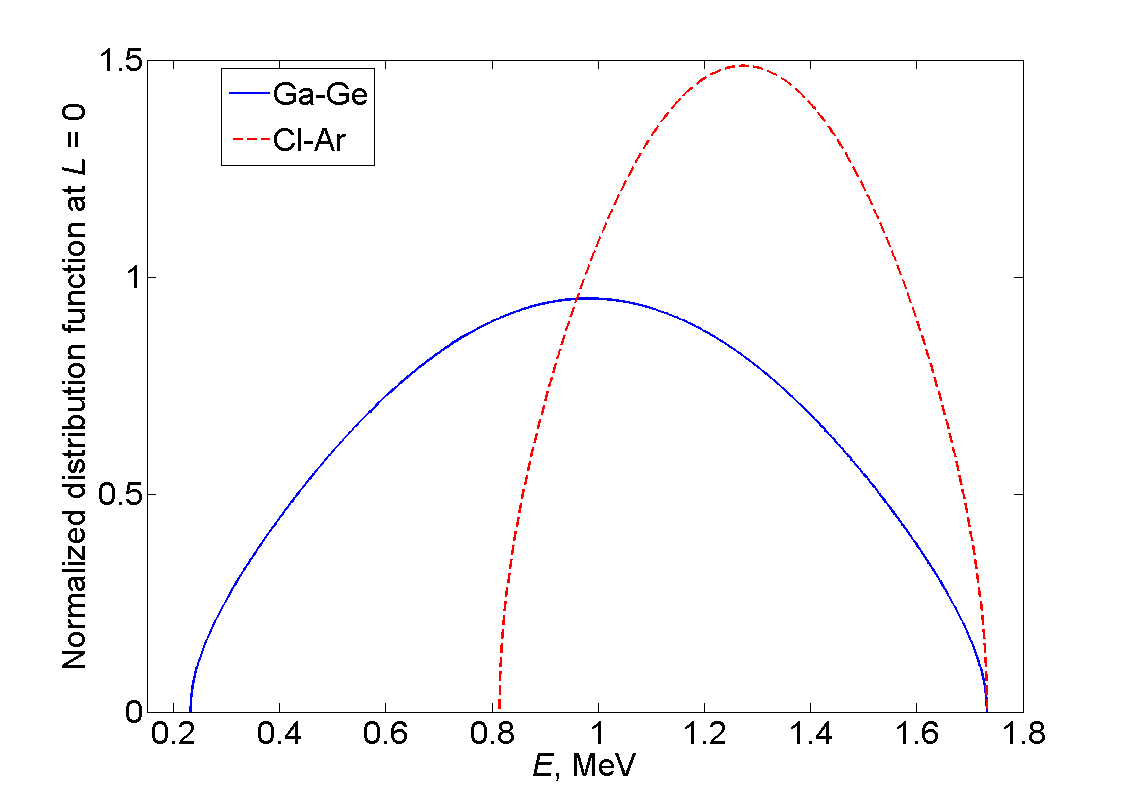}}
\caption{Normalized distribution functions (\ref{distr_1}) for a ${^{15} {\rm O}}$ source and Cl-Ar and Ga-Ge detectors.}
\label{fig_distr_1}
\end{figure}

 Here, again, $\left| {\vec p}
\right|_{\max }$ is determined by the production process and
$\left| {\vec p} \right|_{\min }$ is determined by the detection
process; the explicit expression for the normalization constant
$C$, which is different for different production and registration
processes, is unimportant for us, because we will normalize the
probability (\ref{prob_osc_fin}) so that it equals unity at the
point $L=0$. Normalized distribution function (\ref{distr_1})
represents the relative contribution of the neutrinos with a given
momentum to the probability of the whole process at $L=0$. For the
production and detection processes under consideration we have:
$$\left| {\vec p} \right|_{\min }^\textit{\rm Ga-Ge}  = 232 {\ \rm keV},
\quad \left| {\vec p} \right|_{\min }^\textit{\rm Cl-Ar}  = 814 {\
\rm keV}, \quad \left| {\vec p} \right|_{\max }^\textit{\rm O} =
1732 {\ \rm keV.}$$ Functions (\ref{distr_1}) for  both detectors
are depicted in Fig.~\ref{fig_distr_1}.

\begin{figure}[htb]
\begin{minipage}[h]{0.49\linewidth}
\begin{center}
\includegraphics[width=1\linewidth]{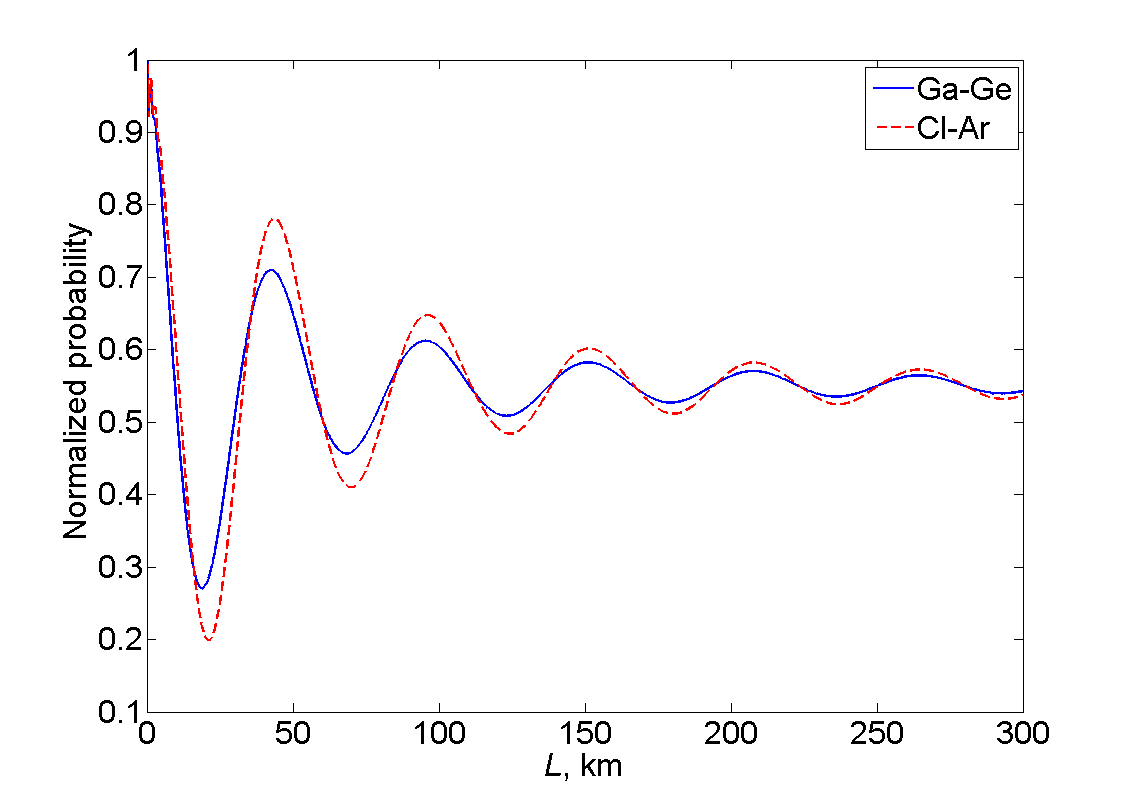} \\ a) Distance $L$ from 0 to 300 km.
\end{center}
\end{minipage}
\hfill
\begin{minipage}[h]{0.49\linewidth}
\begin{center}
\includegraphics[width=1\linewidth]{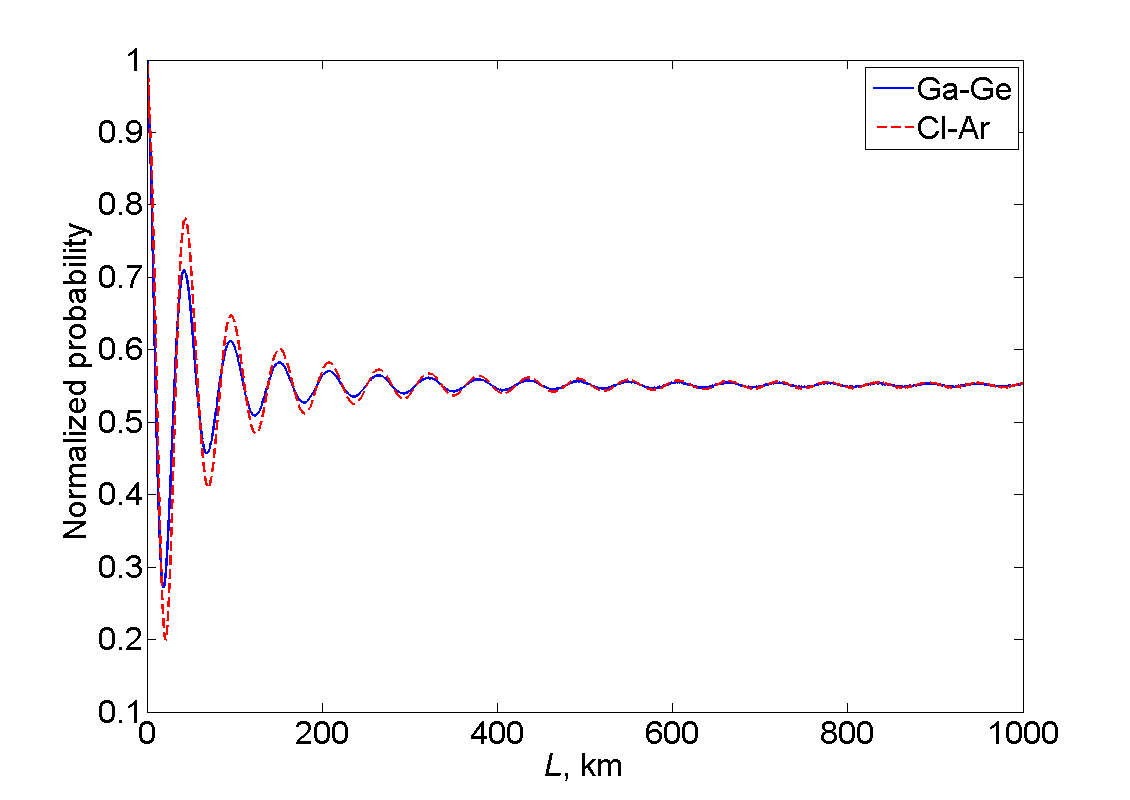} \\ b) Distance $L$ from 0 to 1000 km.
\end{center}
\end{minipage}
\caption{Normalized probabilities of the neutrino oscillation
processes with  the neutrino production in the ${^{15} {\rm O}}$
decay and the registration by Cl-Ar and Ga-Ge detectors.}
\label{fig_osc_1}
\end{figure}

 Below the following values of the neutrino masses and mixing
angles are used \cite{Tanabashi:2018oca}:
$$m_2^2 - m_1^2  = 7.53 \cdot 10^{ - 5} {\ \rm eV}^2, \quad m_3^2  - m_2^2  = 2.51 \cdot 10^{ - 3} {\ \rm eV}^2,$$
$$\theta _{12}  = 0.587, \quad \theta _{13}  = 0.146, \quad \theta _{23}  = 0.702.$$
 We failed to perform the integration in formula
(\ref{prob_osc_fin}) with  probability density (\ref{distr_1})
analytically.  The results of numerical integration are presented
in Fig.~\ref{fig_osc_1} (the probability is normalized to its
value at the point $L=0$).
We see that the oscillation pattern depends on the detection
process and  the oscillations fade out with distance, which gives
rise to a coherence length in our approach. This is due to the
momentum distribution of the intermediate neutrinos. By analogy
with interference in optics we introduce the visibility function:
\begin{equation}
V\left( L \right) = \frac{{I_{\max }  - I_{\min } }}{{I_{\max }  + I_{\min } }}.
\end{equation}
Here $I_{\max}$, $I_{\min}$ stand for the relative neutrino
registration probabilities in the adjacent maximum and minimum of
the oscillation pattern. If we assume the condition of
oscillations' visibility to be $V(L) > 0.1$ (which is standard in
optics), we arrive at the coherence lengths
$$L_{{\rm coh}}^\textit{\rm Ga-Ge}  \approx 105 {\rm \ km}, \quad L_{{\rm coh}}^\textit{\rm Cl-Ar}  \approx 146 {\rm \ km}.$$
In the Ga-Ge case we have a wider momentum distribution  than in
the Cl-Ar one, hence the Ga-Ge oscillation fade out more rapidly
thus having a smaller coherence length.

As one can see in Fig.~\ref{fig_osc_1} the oscillations
asymptotically approach the value close to 0.55. The behavior of
the oscillations at large distances, much more than the coherence
length, is in fact determined by  oscillations' average with
respect to the distance $L$. Thus, the asymptotic behavior  of the
oscillation  is given here, according to (\ref{dif_prob_osc}) and
(\ref{P_ee}), by the expression
\begin{equation}
\overline {P_{ee} }  = 1 - 4\sum\limits_{\scriptstyle i,k = 1 \hfill \atop
  \scriptstyle i < k \hfill}^3 {\left| {U_{1i} } \right|^2 \left| {U_{1k} } \right|^2 \frac{1}{2}}  = \sum\limits_{i = 1}^3 {\left| {U_{1i} } \right|^4 },
\end{equation}
which approximately equals to 0.5511 for the taken values of the
mixing angles $\theta_{ik}$.

\begin{figure}[htb]
\center{\includegraphics[width=0.5\linewidth]{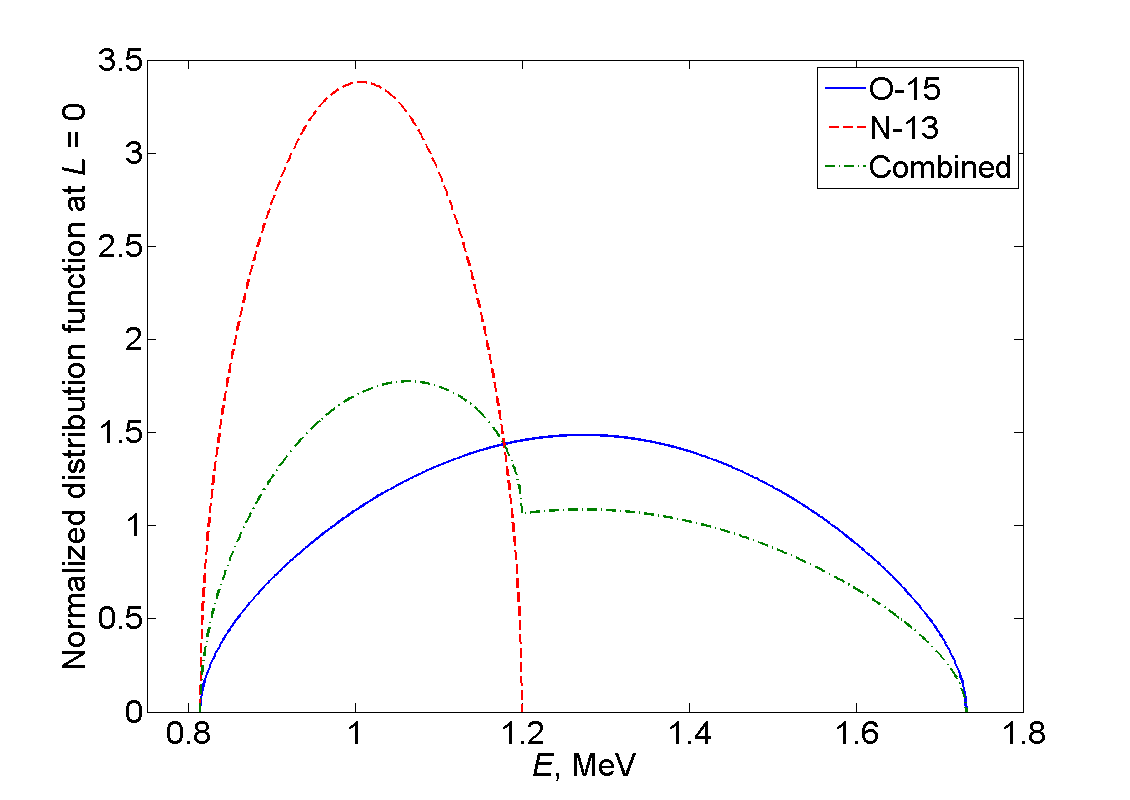}}
\caption{Normalized distribution functions (\ref{distr_1}) for ${^{15} {\rm O}}$, ${^{13} {\rm N}}$ and combined sources and a Cl-Ar detector.}
\label{fig_distr_2}
\end{figure}

\begin{figure}[htb]
\begin{minipage}[h]{0.49\linewidth}
\begin{center}
\includegraphics[width=1\linewidth]{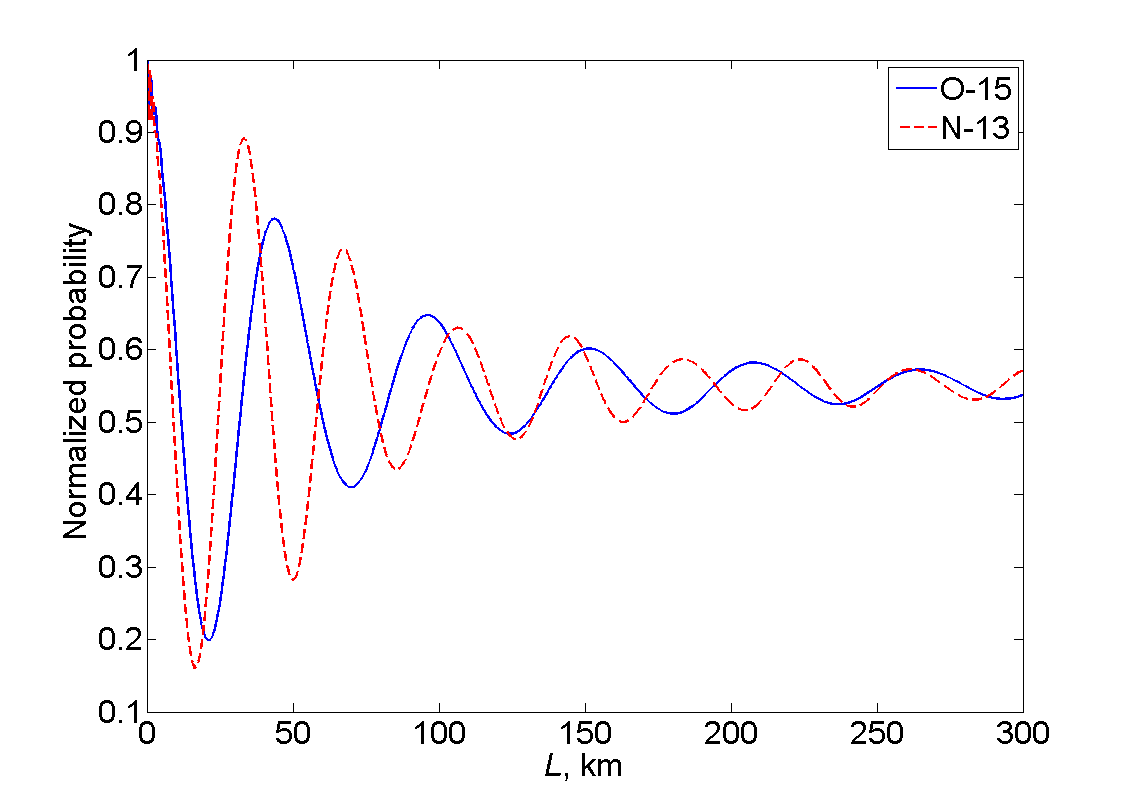} \\ a) Distance $L$ from 0 to 300 km.
\end{center}
\end{minipage}
\hfill
\begin{minipage}[h]{0.49\linewidth}
\begin{center}
\includegraphics[width=1\linewidth]{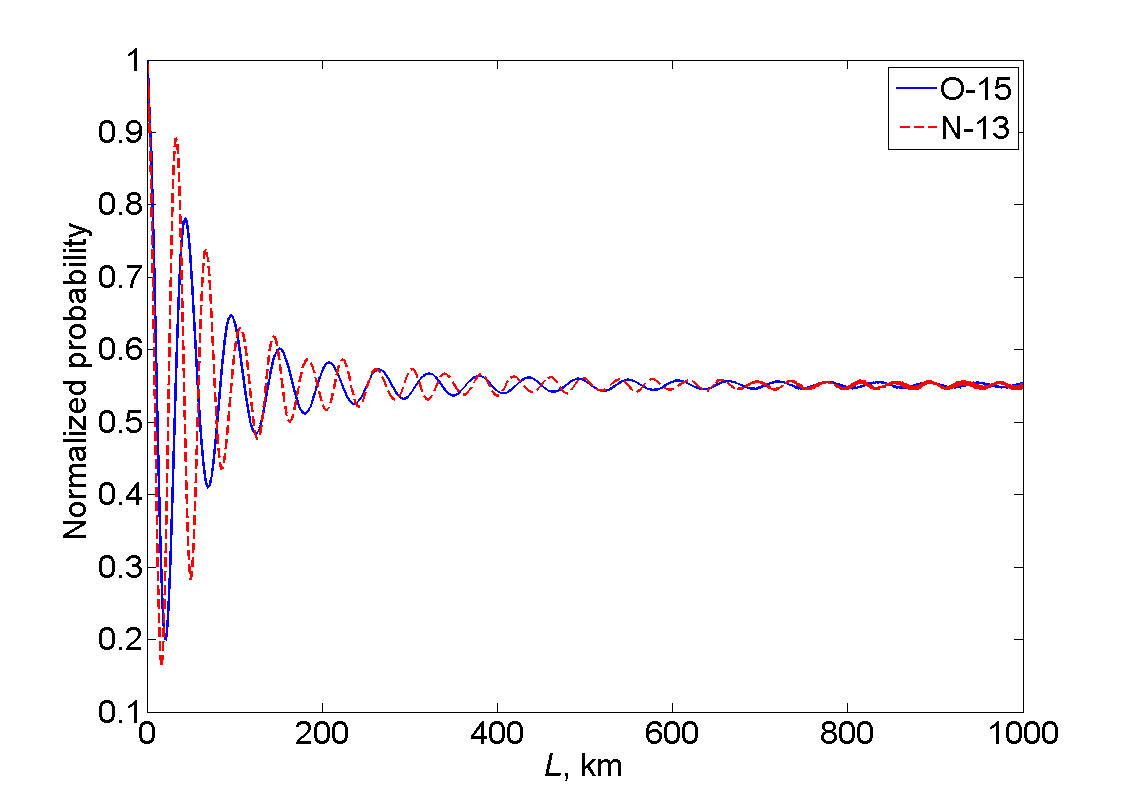} \\ b) Distance $L$ from 0 to 1000 km.
\end{center}
\end{minipage}
\caption{Normalized probabilities of the neutrino oscillation
processes  with the neutrino production in the ${^{15} {\rm O}}$
and ${^{13} {\rm N}}$ decays and the registration by a Cl-Ar
detector.} \label{fig_osc_2}
\end{figure}

Our next step is to compare the neutrino oscillation processes,
where the neutrinos are produced in the  reactions of the solar
carbon cycle
$${^{15} {\rm O}} \to {^{15} {\rm N}} + e^ +   + \nu_i \quad {\rm or} \quad {^{13} {\rm N}} \to {^{13} {\rm C}} + e^ +   + \nu_i $$
and are registered  in a chlorine-argon detector.  For the ${^{13}
{\rm N}}$ decay we have $\left| {\vec p} \right|_{\max
}^\textit{\rm N}  = 1199 {\ \rm keV}$. Normalized functions
(\ref{distr_1}) for these two cases are presented in Fig.~\ref{fig_distr_2} (solid and dashed lines).
The results of the numerical integration with the same
parameters are shown in Fig.~\ref{fig_osc_2}.
The coherence length for the ${^{13} {\rm N}}$ source turns out to be
$$L_{{\rm coh}}^\textit{\rm N}  \approx 158 {\rm \ km,}$$
which is larger than for the previously found ${^{15} {\rm O}}$
case (146 km) since the ${^{13} {\rm N}}$ source provides a more
narrow neutrino momentum distribution.

Finally let us consider a more realistic combined source,  where
the neutrinos are produced in the ${^{15} {\rm O}}$ and ${^{13}
{\rm N}}$ decays simultaneously. The registration is performed
again by a chlorine-argon detector. When a neutrino is detected,
one cannot distinguish, whether it came from a ${^{15} {\rm O}}$
or ${^{13} {\rm N}}$ nucleus. The calculations show that if the
source is in the state of dynamic equilibrium, the probability of
a neutrino being produced by a ${^{13} {\rm N}}$ decay is
approximately 83\% versus 17\% for an ${^{15} {\rm O}}$ one. We
will sum the probabilities  for ${^{15} {\rm O}}$ and ${^{13} {\rm
N}}$ given by formula (\ref{prob_osc_fin}) with different weights,
and these  probabilities of a neutrino being produced in one of
two decays are one source of the weights.

Another source is as follows.  Function (\ref{distr_1}) includes
the constant $C$, which is different for our two cases. Let us
introduce the notations $C_\textit{\rm O}$ and $C_\textit{\rm N}$
for the corresponding coefficients. In our approximation, these
constants satisfy the relations
\begin{equation}\label{C}
4 \pi\int\limits_0^{\left| {\vec p} \right|_{\max }^Z } {C_Z \sqrt
{\left( {\left| {\vec p} \right|_{\max }^Z  - \left| {\vec p}
\right|} \right)\left( {\left| {\vec p} \right|_{\max }^Z  -
\left| {\vec p} \right| + 2m} \right)} \left( {\left| {\vec p}
\right|_{\max }^Z  - \left| {\vec p} \right| + m} \right)\left|
{\vec p} \right|^2 d\left| {\vec p} \right|}  =
\frac{{{\mathop{1}\nolimits} }}{{\tau _Z }},
\end{equation}
where the index $Z$ takes values ``O'' or ``N'' and $\tau _Z$ is
the lifetime of the corresponding nucleus. Given that $\tau
_\textit{\rm O} = 122.24$ sec and $\tau _\textit{\rm N} = 597.90$
sec, performing the numerical evaluation of the integral in
(\ref{C}), one finds the ratio of the coefficients to be
${{C_\textit{\rm O} } \mathord{\left/
 {\vphantom {{C_\textit{\rm O} } {C_\textit{\rm N} }}} \right.
 \kern-\nulldelimiterspace} {C_\textit{\rm N} }} = 1.0548$, which gives a small correction.

\begin{figure}[htbp]
\begin{minipage}[h]{0.49\linewidth}
\begin{center}
\includegraphics[width=1\linewidth]{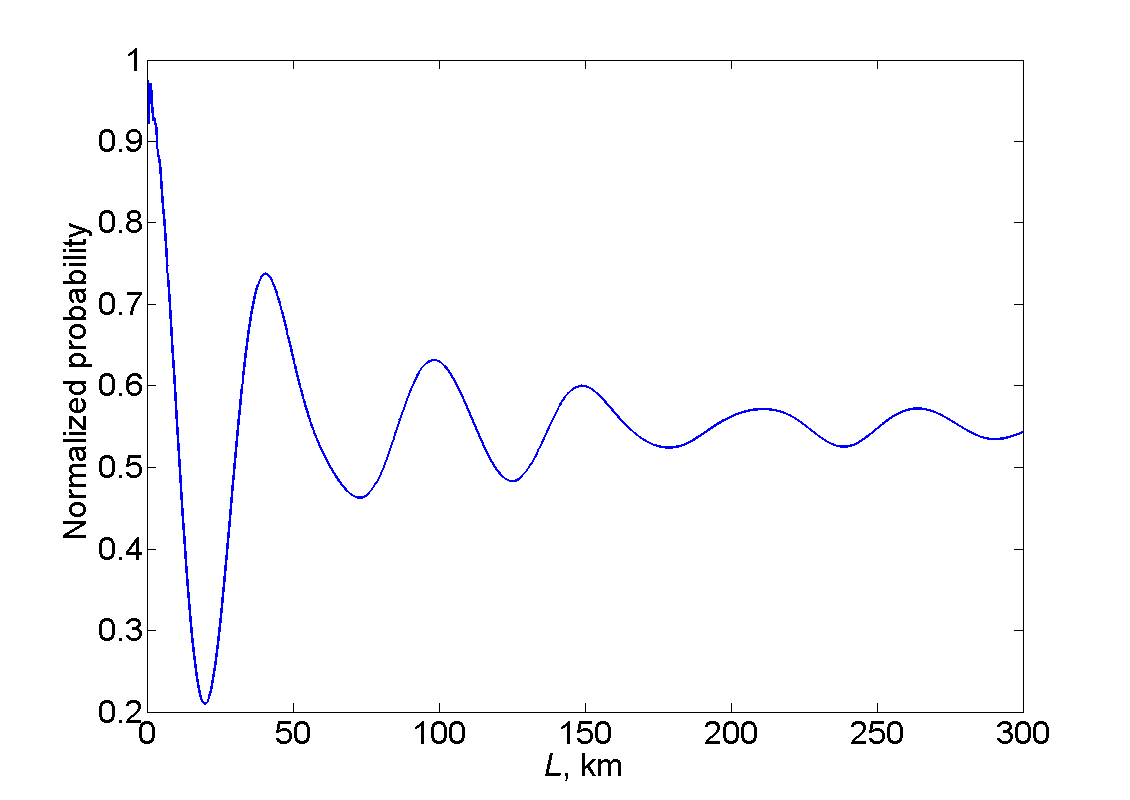} \\ a) Distance $L$ from 0 to 300 km.
\end{center}
\end{minipage}
\hfill
\begin{minipage}[h]{0.49\linewidth}
\begin{center}
\includegraphics[width=1\linewidth]{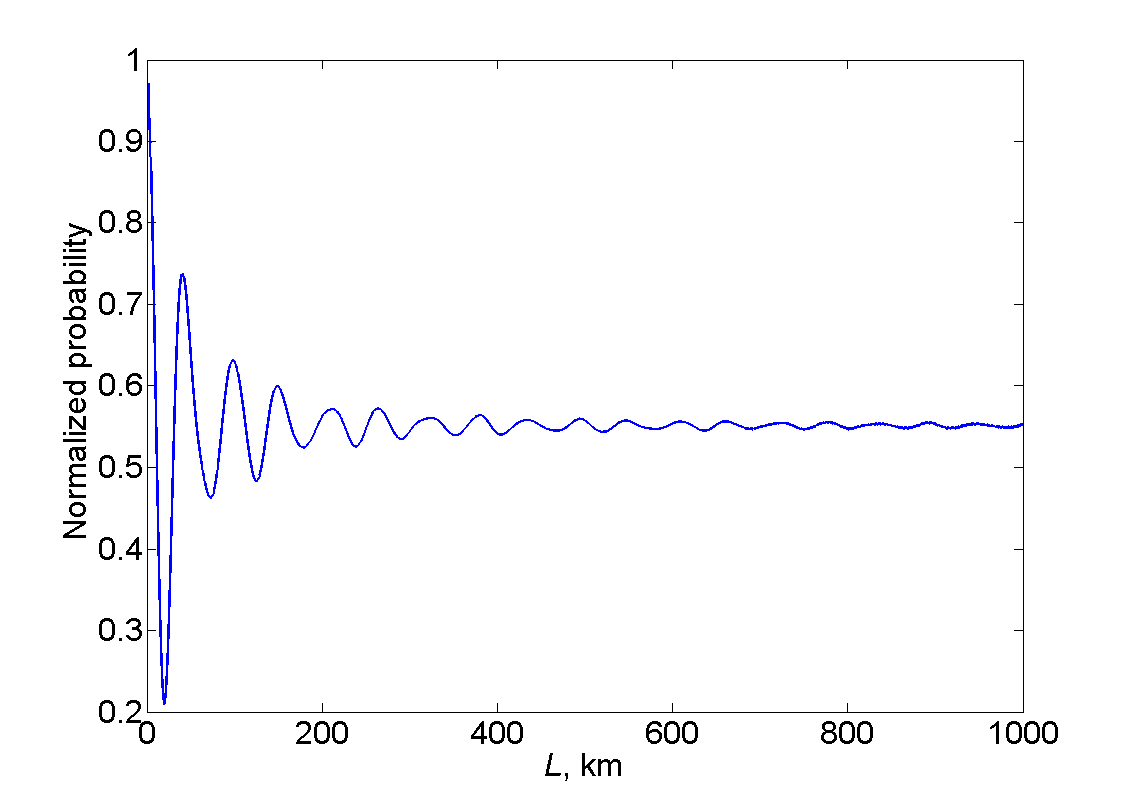} \\ b) Distance $L$ from 0 to 1000 km.
\end{center}
\end{minipage}
\caption{Normalized probability of the neutrino oscillation process with the neutrino production in both the ${^{15} {\rm O}}$ and ${^{13} {\rm N}}$ decays simultaneously and the registration by a Cl-Ar detector.}
\label{fig_osc_3}
\end{figure}

The resulting weights of  probabilities (\ref{prob_osc_fin}) for
the ${^{15} {\rm O}}$ and ${^{13} {\rm N}}$ nuclei are the
products of the corresponding coefficients from these two sources.
The weights can
be chosen in a transparent way to be 0.1776 for the ${^{15} {\rm
O}}$ contribution and 0.8224 for the ${^{13} {\rm N}}$ contribution.
Total normalized function (\ref{distr_1}) for such an experiment
is presented in Fig.~\ref{fig_distr_2} (dash-dotted line). The results
of the numerical integration are depicted in Fig.~\ref{fig_osc_3}.
The overlapping of the oscillation patterns from two different
sources  leads to an even more rapid  fading out of the
oscillations, and in this case the coherence length reads
$$L_{{\rm coh}}^\textit{\rm O+N}  \approx 142 {\rm \ km,}$$
which is less than for the ${^{15} {\rm O}}$ or ${^{13} {\rm N}}$
sources separately.

At the end of this section we would like to stress ones again that
the coherence length discussed above differs essentially from the
coherence length appearing in the standard quantum-mechanical
description of neutrino oscillation in terms of wave packets. In
the latter  case the oscillation fading out and, consequently, the
coherence length arises due to the quantum-mechanical uncertainty
of  neutrino momentum. In contrast to it, in the approach under
consideration the neutrinos are supposed to have definite momenta
(no momentum uncertainty), and the  origin of the oscillation
pattern blurring is the  momentum distribution of the intermediate
neutrinos. It is always present in a three-body decay, even if all
the initial and final particles and nuclei have definite momenta.
The above calculations show that this cause of oscillation fading
out leads to  much smaller coherence lengths than the ones, which
are due to the natural momentum uncertainty  considered in the
standard approach. It means that the effect of neutrino
non-monochromaticity taken into account in the framework of our
approach is dominant in a realistic experimental setting, while
the blurring due to the neutrino momentum uncertainty can be
neglected compared to it.

\section{Neutrino oscillations in experiments with detection in the charged-current interaction only}

\subsection{Theory}

In the same way one can consider the neutrino oscillation process,
where the neutrinos  are produces in the charged-current
interaction with nuclei and detected in both the charged- and
neutral-current interactions with an electron. The process is
described by the following diagrams:
\vspace{1.5cm}
\begin{figure}[h]
\begin{center}
$\qquad \quad \ \! \! \! \: \! \: \! $ \includegraphics[width=0.4434\linewidth]{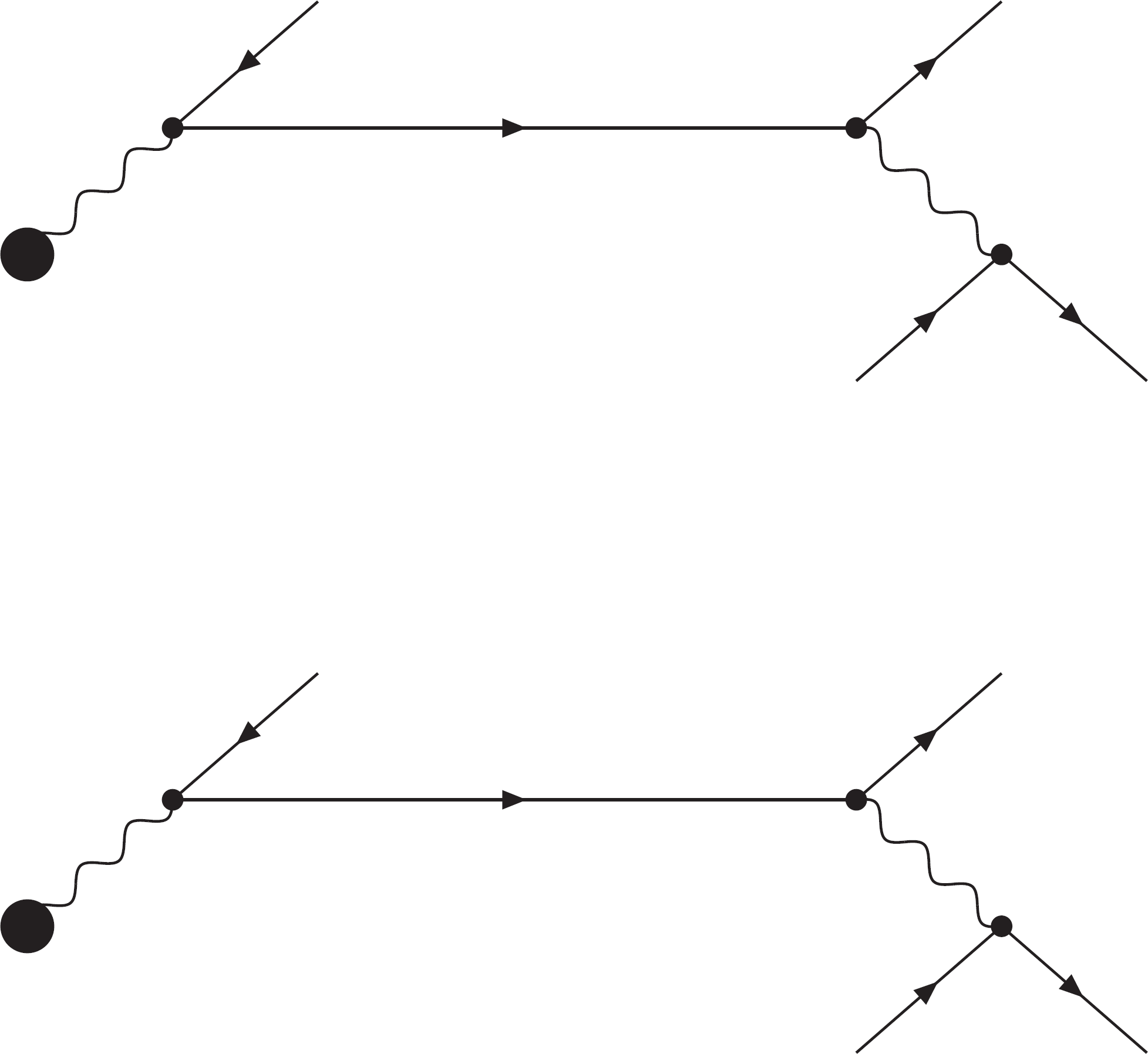}
\end{center}
\end{figure}
\vspace*{-7.967cm}
\begin{center}
\begin{picture}(193,81)(0,0)
\Text(70.0,94.0)[l]{$e^+ ( q)$}\ArrowLine(67.5,88.0)(40.5,64.5)
\Text(33.5,65.5)[r]{$x$} \Photon(13.5,41.0)(40.5,64.5){2}{3.0}
\Text(53.5,48.5)[r]{$W^+$} \Vertex (13.5,41.0){5} \Vertex
(40.5,64.5){2} \ArrowLine(40.5,64.5)(167.5,64.5) \Vertex
(167.5,64.5){2} \Text(104.8,70.5)[b]{$\nu_i ( p_{\rm n} )$}
\ArrowLine(167.5,64.5)(194.5,88.0) \Text(198.0,94.0)[l]{$\nu_i ( k_2 )$}
\Text(175.0,64.5)[l]{$y$} \Text(166.0,48.5)[l]{$Z$}
\Photon(167.5,64.5)(194.5,41.0){2}{3.0} \Vertex (194.5,41.0){2}
\ArrowLine(167.5,17.5)(194.5,41.0) \Text(132.5,17.5)[l]{$e^- ( k_1 )$}
\ArrowLine(194.5,41.0)(221.5,17.5) \Text(225.0,17.5)[l]{$e^- ( k )$}
\Text(330.0,60.5)[b]{\addtocounter{equation}{1}(\arabic{equation})}
\setcounter{diag3}{\value{equation}}
\end{picture}
\end{center}
\vspace{0.7cm}
\begin{center}
\begin{picture}(193,81)(0,0)
\Text(70.0,94.0)[l]{$e^+ (q)$}\ArrowLine(67.5,88.0)(40.5,64.5)
\Text(33.5,65.5)[r]{$x$} \Photon(13.5,41.0)(40.5,64.5){2}{3.0}
\Text(53.5,48.5)[r]{$W^+$} \Vertex (13.5,41.0){5} \Vertex
(40.5,64.5){2} \ArrowLine(40.5,64.5)(167.5,64.5) \Vertex
(167.5,64.5){2} \Text(104.8,70.5)[b]{$\nu_k ( p_{\rm n} )$}
\ArrowLine(167.5,64.5)(194.5,88.0) \Text(197.5,94.0)[l]{$e^- ( k )$}
\Text(175.0,64.5)[l]{$y$} \Text(160.0,48.5)[l]{$W^+$}
\Photon(167.5,64.5)(194.5,41.0){2}{3.0} \Vertex (194.5,41.0){2}
\ArrowLine(167.5,17.5)(194.5,41.0) \Text(132.5,17.5)[l]{$e^- ( k_1 )$}
\ArrowLine(194.5,41.0)(221.5,17.5) \Text(225.0,17.5)[l]{$\nu_i ( k_2 )$}
\Text(330.0,60.5)[b]{\addtocounter{equation}{1}(\arabic{equation})}
\setcounter{diag4}{\value{equation}}
\end{picture}
\end{center}
The amplitude corresponding to diagram (\arabic{diag4}) should be
summed over all the three neutrino mass eigenstates, i.e.\
$k=1,2,3$, as they all contribute. Since only the final electron
is detected in the experiment, the probability of the process with
$i$-th neutrino mass eigenstate in the final state should be
summed over $i$ to give us the probability of registering an
electron.

Now let us denote the particle momenta as follows: the momentum of
the positron is $q$, the momentum of the virtual neutrinos is
$p_{\rm n}$, the momentum of the outgoing electron is $k$, the
momentum of the incoming electron is $k_1$, the momentum of the
outgoing neutrino is $k_2$, the momentum of the initial nucleus is
$P^{(1)} = \left( { E^{(1)}, \vec P^{(1)} } \right)$ and the
momentum of the final nucleus is $P^{(1^\prime)} = \left( {
E^{(1^\prime)}, \vec P^{(1^\prime)} } \right)$ (we retain the
notations of the previous section for the nuclear values in order
to use the formulas from it without redefinitions).

Again we use the approximation of Fermi's interaction and take the
time-dependent propagator (\ref{spin_prop_1}) keeping the neutrino
masses only in the exponential. The amplitude corresponding to
diagram (\arabic{diag3}) in the momentum representation, when $y^0
- x^0 = T$, looks like
\begin{eqnarray}
\label{amp_nc} M_{\rm nc}^{(i)} &=& i \frac{{G_{\rm F}^{\,2} }}{4p_{\rm n}^0}\, { U_{1i} ^* } e^{
- i\frac{{m_i^2  - p_{\rm n}^2 }}{{2 { p^0_{\rm n}} }}T}\, \bar \nu_i
\left( k_2 \right)\gamma ^\mu \left( {1 - \gamma ^5 } \right) \hat
p_{\rm n}\gamma ^\rho \left(
{1 - \gamma ^5 } \right) v \left( q \right)j_\rho^{(1)} \left( {P^{(1)}, P^{(1^\prime)} } \right) \times \\
 & & \times \left[ \left( -\frac{1}{2} + \sin^2 \theta_{\rm W} \right)
\bar u \left( k \right)\gamma _\mu (1 - \gamma^5) u \left( k_1
\right) + \sin^2 \theta_{\rm W} \bar u \left( k \right)\gamma _\mu (1 +
\gamma^5) u \left( k_1 \right)\right]. \nonumber
\end{eqnarray}
Similarly, the amplitude corresponding to diagram (\arabic{diag4}) summed over $k$ reads
\begin{eqnarray}
M_{\rm cc}^{(i)} & = &-i \frac{{G_{\rm F}^{\,2} }}{4{p_{\rm n}^0}}\, {U_{1i}^*
}\left( \sum\limits_{k = 1}^3 {\left| {U_{1k} } \right|^2} e^{ -
i\frac{{m_i^2  - p_{\rm n}^2 }}{{2 { p^0_{\rm n}} }}T} \right) \bar u
\left( k \right)\gamma ^\mu (1 - \gamma^5) \hat p_{\rm n}\gamma ^\rho
\left( {1 - \gamma ^5 } \right)
v\left( q \right) \times \nonumber \\
\label{amp_cc} & & \times j_\rho^{(1)} \left( {P^{(1)}, P^{(1^\prime)} } \right) \bar \nu_i \left(
k_2 \right)\gamma _\mu (1 - \gamma^5) u \left( k_1 \right).
\end{eqnarray}

The squared modulus of the total amplitude $M_{\rm tot}^{(i)}  =
M_{\rm nc}^{(i)}  + M_{\rm cc}^{(i)}$,  averaged with respect and summed
over particles' polarizations, factorizes in the approximation
$p_{\rm n}^2=0$ as follows:
\begin{equation} \label{sqr_amp_nc}
\left\langle {\left| M_{\rm tot}^{(i)} \right|^2 } \right\rangle  =
\left\langle {\left| M_1 \right|^2 } \right\rangle \left\langle {\left| M_2^{(i)} \right|^2 } \right\rangle \frac{{1}}{{4 (p_{\rm n}^0)^2 }}.
\end{equation}
Here $\left\langle {\left| M_1 \right|^2 } \right\rangle$ is given by (\ref{sqr_M1}),
\begin{eqnarray}
\left\langle {\left| M_2^{(i)} \right|^2 } \right\rangle &=& 64 G_{\rm F}^{\,2}\left[\left|B_i + A_i \left(-\frac{1}{2} +
\sin^2 \theta_{\rm W} \right)\right|^2 (p_{\rm n}k_1)^2 + \left|A_i\right|^2 \sin^4 \theta_{\rm W}  (p_{\rm n} k)^2 - \right. \nonumber \\
\label{factor2_nc} & & - \sin^2 \theta_{\rm W} \, {\rm Re} \left( \left(B_i + A_i \left(-\frac{1}{2} + \sin^2 \theta_{\rm W} \right)\right) A_i^* \right) (p_{\rm n} k_2)m^2 \Bigg],
\end{eqnarray}
where the notations
\begin{equation}
A_i = { U_{1i}^* } e^{ - i\frac{{m_i^2  - p_{\rm n}^2 }}{{2 { p^0_{\rm n}} }}T}, \quad B_i =
{U_{1i}^* }\left( \sum\limits_{k = 1}^3 {\left| {U_{1k} } \right|^2} e^{ - i\frac{{m_i^2  - p_{\rm n}^2 }}{{2 {p^0_{\rm n}} }}T} \right)
\end{equation}
are introduced.

Following the outlined procedure, we introduce the virtual
neutrino 4-momentum $p$ in the same manner, multiply the squared
amplitude (\ref{sqr_amp_nc}) by the delta function of
energy-momentum conservation $(2\pi )^4 \delta (P^{(1)} + k_1 -
P^{(1^\prime)} - q - k - k_2)$, substitute $p$ instead of $p_{\rm
n}$, multiply by $2\pi \delta ( P^{(1)} - P^{(1^\prime)} - q - p
)$ and integrate the result with respect to the phase volume of
the final particles and nucleus. Next we sum the resulting
differential probability of the process over the final neutrino
type $i$, substitute $T = {{Lp^0 } \mathord{\left/
 {\vphantom {{Lp^0 } {\left| {\vec p} \right|}}} \right.
 \kern-\nulldelimiterspace} {\left| {\vec p} \right|}}$, multiply
 the result by $\left| {\vec p} \right|^2$ and integrate it with
 respect to $\left| {\vec p} \right|$ from $\left| {\vec p} \right|_{\min }$ to $\left| {\vec p} \right|_{\max }$.
 We arrive at the probability of detecting an electron:
\begin{equation}\label{prob_osc_nc_fin}
\frac{{dW}}{{d\Omega }} = \int\limits_{\left| {\vec p} \right|_{\min } }^{\left| {\vec p} \right|_{\max } }
{\frac{{d^3W}}{{d^3 p}}\left| {\vec p} \right|^2 d\left| {\vec p} \right|}  =
\int\limits_{\left| {\vec p} \right|_{\min } }^{\left| {\vec p} \right|_{\max } }
{\frac{{d^3W_1 }}{{d^3 p}}W_2 \left| {\vec p} \right|^2 d\left| {\vec p} \right|} .
\end{equation}
Here $\frac{{d^3W}}{{d^3 p}}$ is the differential probability  of
the whole process, where the intermediate neutrinos have a
definite momentum $\vec p$ and the final neutrino mass eigenstate
is of any type, $\frac{{d^3W_1 }}{{d^3 p}}$ is the differential
probability of decay of the initial nucleus into the final
nucleus, a positron and a massless fermion with the momentum $\vec
p$ given by (\ref{dif_W_1}) and
\begin{eqnarray}
W_2  &=& \frac{1}{{2p^0 2k_1^0 }}\int {\frac{{d^3 k}}{{\left( {2\pi } \right)^3 2k^0 }}
\frac{{d^3 k_2 }}{{\left( {2\pi } \right)^3 2k_2^0 }}\sum\limits_{i = 1}^3
{\left. {\left\langle {\left| {M_2^{(i)} } \right|^2 } \right\rangle } \right|_{\scriptstyle p_{\rm n}  = p \hfill \atop
  \scriptstyle T = {{Lp^0 } \mathord{\left/
 {\vphantom {{Lp^0 } {\left| {\vec p} \right|}}} \right.
 \kern-\nulldelimiterspace} {\left| {\vec p} \right|}} \hfill} } \left( {2\pi } \right)^4 \delta \left( {k_1  + p - k - k_2 } \right)} = \nonumber \\
&=& \frac{{G_{\rm F}^2 m}}{{2\pi }}\frac{{2\left| {\vec p} \right|^2 }}{{2\left| {\vec p} \right| + m}}
\left[ {1 - 2\sin ^2 \theta _{\rm W} \left( {1 + \frac{{2\left| {\vec p} \right|}}{{2\left| {\vec p} \right| + m}}} \right) +
4\sin ^4 \theta _{\rm W} \left( {1 + \frac{1}{3}\left( {\frac{{2\left| {\vec p} \right|}}{{2\left| {\vec p} \right| + m}}} \right)^2 } \right) + } \right. \nonumber \\
 & & \left. { + 4\sin ^2 \theta _{\rm W} \left( {1 + \frac{{2\left| {\vec p} \right|}}{{2\left| {\vec p} \right| + m}}} \right)P_{ee} \left( L \right)} \right] \label{W_2}
\end{eqnarray}
is the probability of the neutrino scattering in the detector. Now
we will use expression (\ref{prob_osc_nc_fin}) to consider several
examples.

\subsection{Specific examples}

In the present subsection we consider neutrino oscillation
experiments, where the neutrinos are produced in the decays of
${^{15} {\rm O}}$ or ${^{13} {\rm N}}$ and registered by a
water-based Cherenkov detector. For  simplicity we assume  that
the final electron is detected, when its speed exceeds the speed
of light in water. It gives us the registration threshold $\left|
{\vec p} \right|_{\min }^\textit{\rm Cher}  = 775 {\ \rm keV}$.

\begin{figure}[htb]
\center{\includegraphics[width=0.5\linewidth]{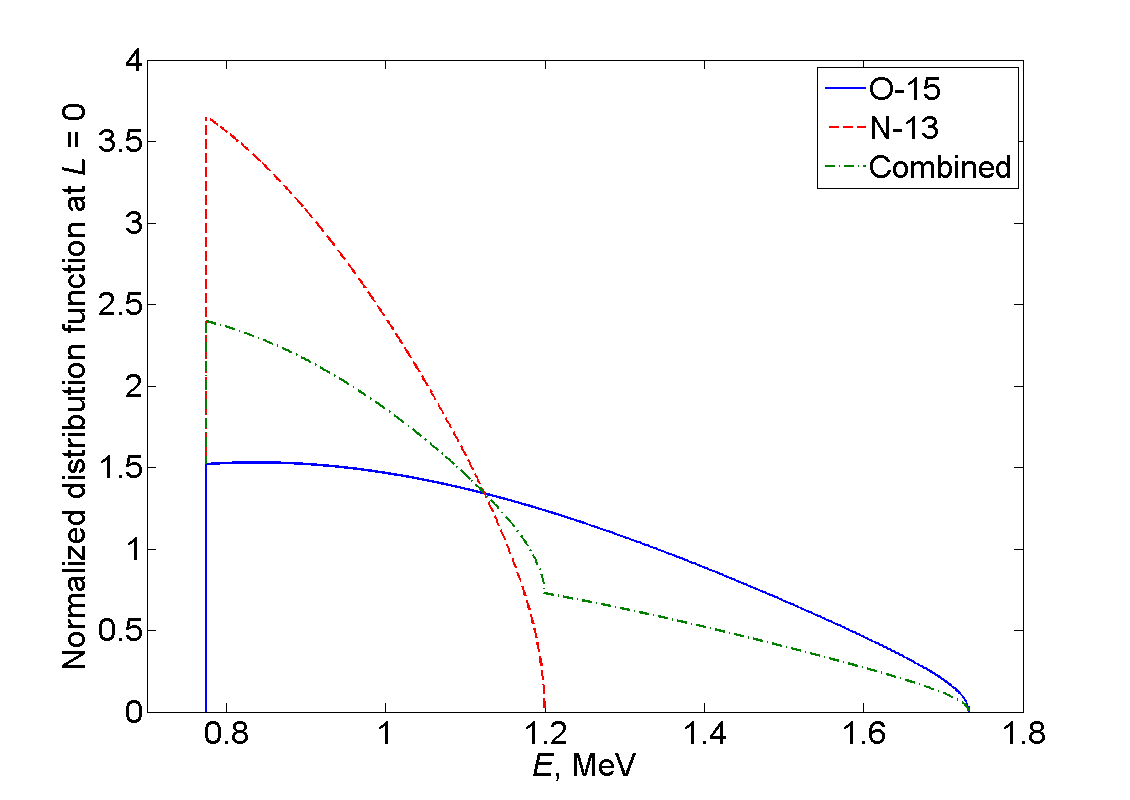}}
\caption{Normalized distribution functions $\frac{{d^3 W_1 }}{{d^3 p}}W_2$ at the point $L=0$ for ${^{15} {\rm O}}$,
${^{13} {\rm N}}$ and combined sources and a water-based Cherenkov
detector.}
\label{fig_distr_3}
\end{figure}

\begin{figure}[htb]
\begin{minipage}[h]{0.49\linewidth}
\begin{center}
\includegraphics[width=1\linewidth]{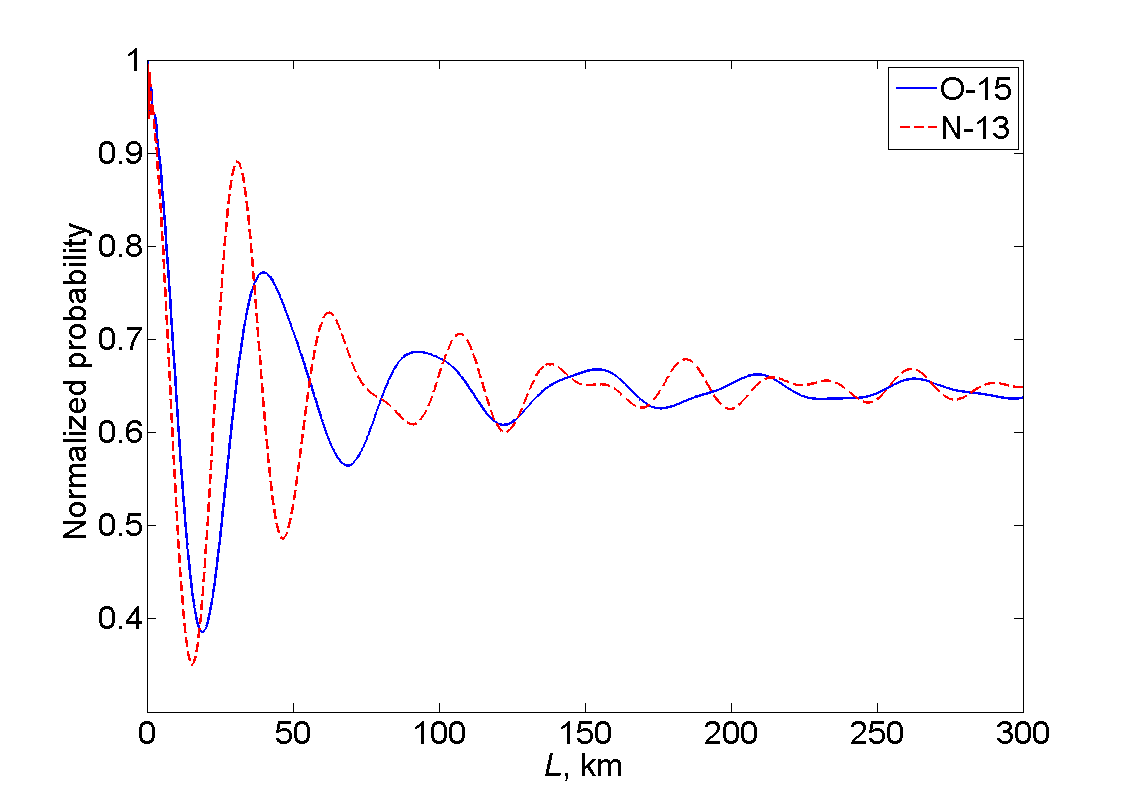} \\ a) Distance $L$ from 0 to 300 km.
\end{center}
\end{minipage}
\hfill
\begin{minipage}[h]{0.49\linewidth}
\begin{center}
\includegraphics[width=1\linewidth]{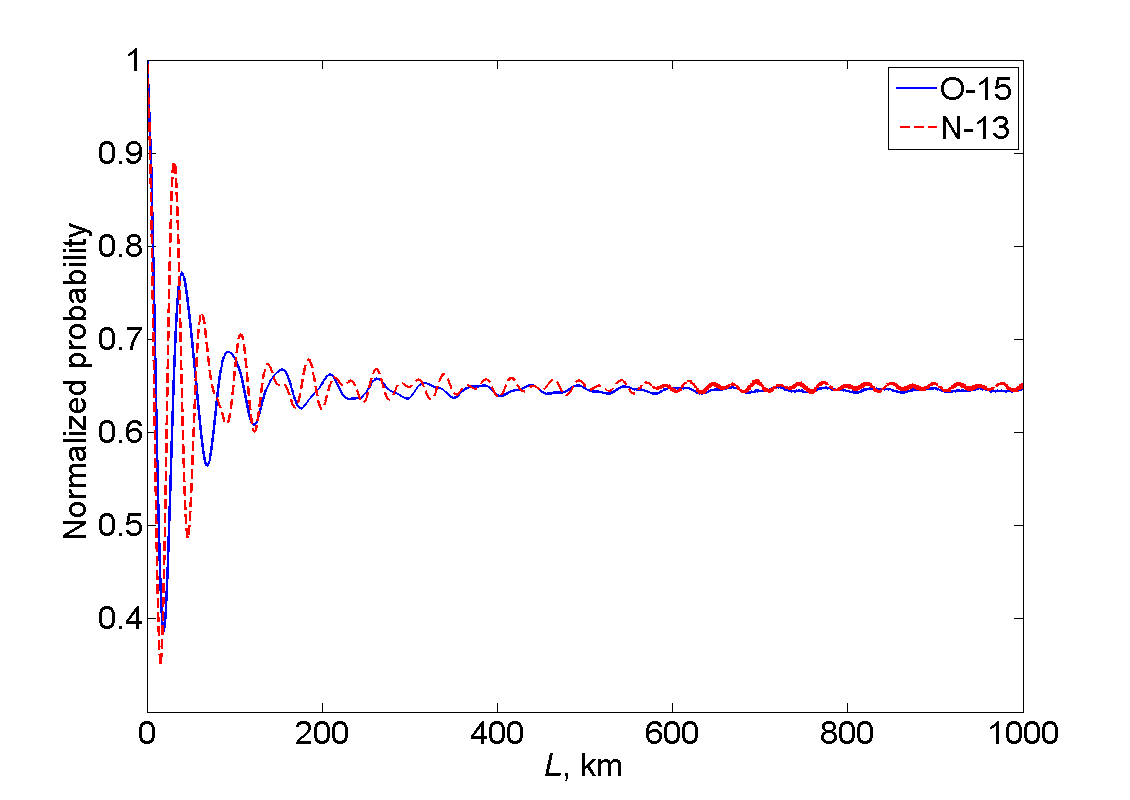} \\ b) Distance $L$ from 0 to 1000 km.
\end{center}
\end{minipage}
\caption{Normalized probabilities of the neutrino oscillation
processes with the neutrino production in the ${^{15} {\rm O}}$
and ${^{13} {\rm N}}$ decays and the registration by a water-based
Cherenkov detector.} \label{fig_osc_4}
\end{figure}

Neglecting the dependence of the  nuclear form-factors on the
momentum transfer, we can again approximate the differential
probability of neutrino production by the function
\begin{equation}
\frac{{d^3 W_1 }}{{d^3 p}} = C\sqrt {\left( {\left| {\vec p} \right|_{\max }  -
\left| {\vec p} \right|} \right)\left( {\left| {\vec p} \right|_{\max }  -
\left| {\vec p} \right| + 2m} \right)} \left( {\left| {\vec p} \right|_{\max }  - \left| {\vec p} \right| + m} \right).
\end{equation}
The normalized distribution functions $\frac{{d^3 W_1 }}{{d^3 p}}W_2$
at the point $L=0$ in this approximation are represented in Fig.~\ref{fig_distr_3} (solid and dashed lines).
The results of numerical integration with the same parameters as were used in the previous section are depicted in Fig.~\ref{fig_osc_4}.
The irregular form of the oscillation pattern in this  case is due
to the sharp cut of the neutrino momentum distribution, defined by
the detection threshold. The coherence lengths here turn out to be
$$L_{{\rm coh}}^\textit{\rm O}  \approx 80 {\rm \ km}, \quad L_{{\rm coh}}^\textit{\rm N}  \approx 75 {\rm \ km.}$$

\begin{figure}[htb]
\begin{minipage}[h]{0.49\linewidth}
\begin{center}
\includegraphics[width=1\linewidth]{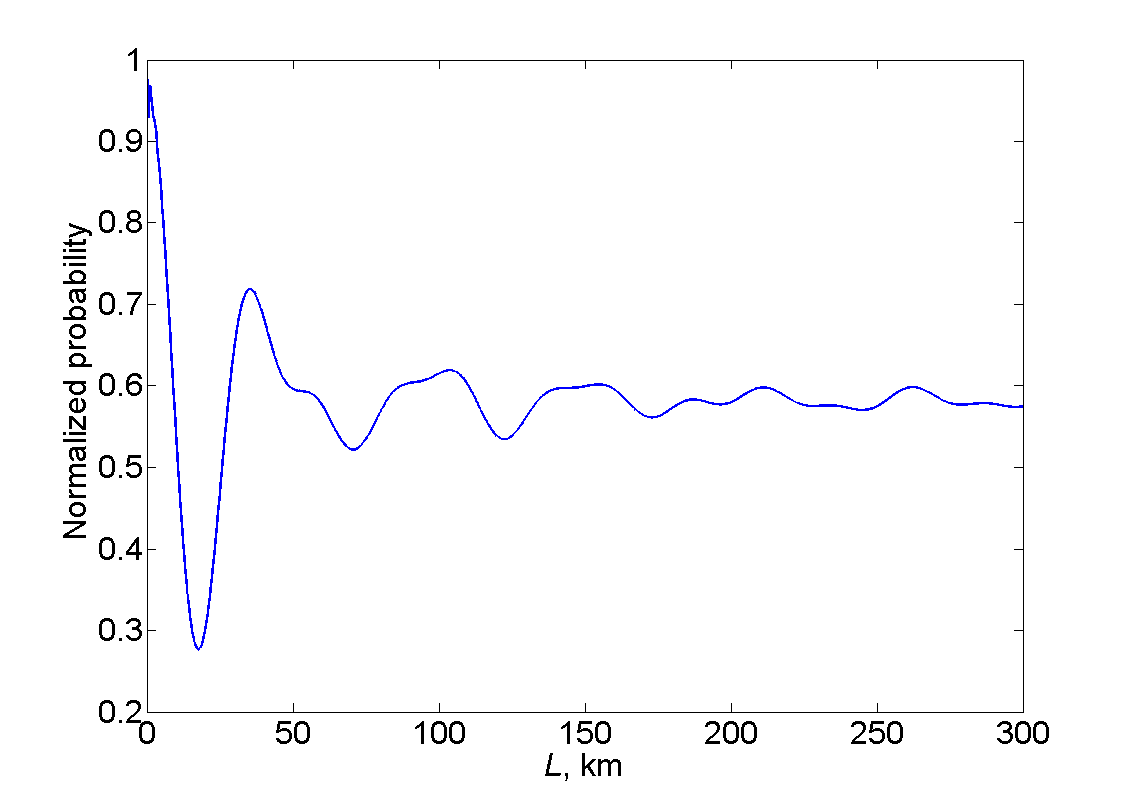} \\ a) Distance $L$ from 0 to 300 km.
\end{center}
\end{minipage}
\hfill
\begin{minipage}[h]{0.49\linewidth}
\begin{center}
\includegraphics[width=1\linewidth]{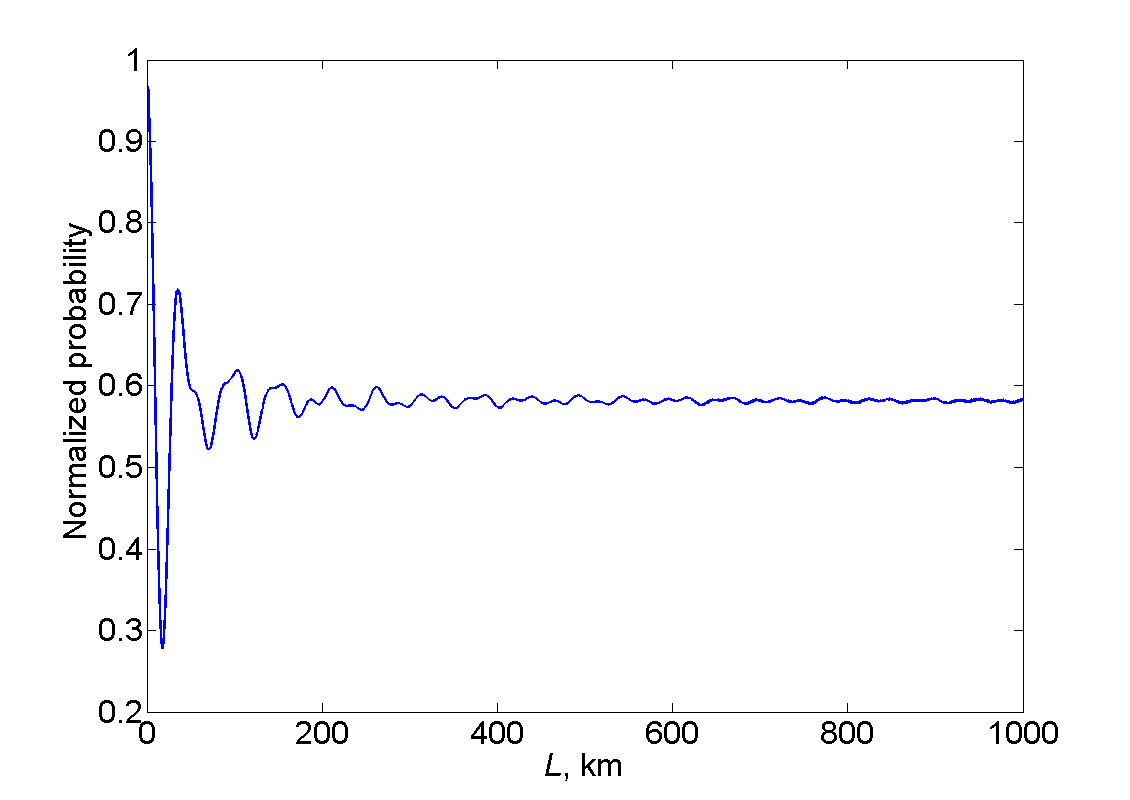} \\ b) Distance $L$ from 0 to 1000 km.
\end{center}
\end{minipage}
\caption{Normalized probability of the neutrino oscillation
process  with the neutrino production in both the ${^{15} {\rm
O}}$ and ${^{13} {\rm N}}$ decays simultaneously and the
registration by a water-based Cherenkov detector.}
\label{fig_osc_5}
\end{figure}

Finally let us consider the combined ${^{15} {\rm O}}$ and ${^{13}
{\rm N}}$ source.  The summation of the probabilities is performed
with the same weights as it was discussed in the previous section.
The normalized distribution function $\frac{{d^3 W_1 }}{{d^3
p}}W_2$ at the point $L=0$ for this case is shown in Fig.~\ref{fig_distr_3} (dash-dotted line). The results of numerical
integration are presented in Fig.~\ref{fig_osc_5}.
The coherence length reads
$$L_{{\rm coh}}^\textit{\rm O+N}  \approx 32 {\rm \ km,}$$
which is, as expected, less than for the ${^{15} {\rm O}}$ and
${^{13} {\rm N}}$ sources separately.

We would like to note here that, unlike the case of registration
only in the charged-current interaction,  discussed in Section 2,
the asymptotic values of the normalized probabilities of the
neutrino oscillation processes presented in Figs.~\ref{fig_osc_4},
\ref{fig_osc_5} are all different. This is due to the fact that,
in the case of registration in the charged-current interaction
only, the oscillating expression $P_{ee} \left( L \right)$, given
by (\ref{P_ee}), factorizes. Thus, the oscillation asymptotic
behavior is determined by the average of $P_{ee} \left( L
\right)$. However, when the registration is performed in both the
charged- and neutral-current interactions, there is no such
factorization, as one can see from formulas
(\ref{prob_osc_nc_fin})--(\ref{W_2}). The numerical evaluation
gives that in the case of ${^{15} {\rm O}}$ and ${^{13} {\rm N}}$
sources separately the asymptotic values are close to each other
and read 0.6454 and 0.6489, respectively, whereas in the case of
combined source  the asymptotic value is 0.5818.

\section{Conclusion}
 A novel quantum field-theoretical approach to the description
of neutrino oscillation processes passing  at finite space-time
intervals is discussed. It is based on the Feynman diagram
technique in the coordinate representation supplemented by
modified rules of passing to the momentum representation, which
reflect the experimental situation at hand. Wave packets are not
employed, we use only the description in terms of plane waves,
which considerably simplifies the calculations. The neutrino
flavor states turn out to be unnecessary and only the neutrino
mass eigenstates are used.

We have explicitly shown that the approach allows one to
consistently  describe the processes of  neutrino oscillations.
The predictions for the probabilities of these processes are found
to completely coincide with the results obtained  in the standard
quantum-mechanical approach.

The approach under consideration also predicts a suppression of
neutrino oscillations with distance. In the standard
quantum-mechanical description this suppression is assumed to
arise due to the quantum uncertainty of neutrino momentum.
However,  in a realistic experiment there is also another source
of the suppression effect. It is the intermediate neutrino being
non-monochromatic, which always takes place in the case of a
tree-body decay even if all the involved particles are assumed to
have definite momenta. If the production process has a
two-particle final state, the momentum spread of the neutrinos
comes from the momentum spread of the initial particles and/or
nuclei. In any realistic experimental situation a neutrino
momentum distribution of this type is always present and is
determined by the spectral characteristics of the production and
detection processes. The width of this distribution is much larger
than that of the natural neutrino momentum distribution due to the
quantum-mechanical uncertainty, considered in the standard
approach. Consequently, the corresponding coherence length turns
out to be much smaller than the one predicted in the standard
quantum-mechanical approach, and hence the former coherence length
is dominant in  experiments. The decoherence process caused by the
neutrino momentum quantum uncertainty also affects the oscillation
pattern blurring, but we can neglect it compared to the more
powerful effect due to  the momentum spread of the intermediate
neutrinos.

In the approach under consideration neutrino oscillation is an
interference process, and  the coherence length  is found by
analogy with interference  of non-monochromatic light in optics
with the help of the visibility function. It is completely defined
by the production and detection processes and cannot be decomposed
into coherence lengths for pairs of neutrino mass eigenstates. The
coherence lengths for five combinations of production and
detection processes have been explicitly calculated. It was found
that the coherence length in the experiments with two production
processes is smaller than the coherence length   in experiments
with only one of the production processes  and with the same
detection process.

It is necessary to mention that, in the developed approach, there
is no analogue of the localization term, which appears in the
wave-packet treatment of neutrino oscillations. This is due to the
fact that the approach under consideration is based on the
assumption that the sizes of the neutrino source and detector are
much smaller than the distance between them, which is always
fulfilled in neutrino oscillation experiments. Since the coherence
length is of the order of the latter distance, this means that the
production and detection processes are localized in space-time
regions much smaller than the oscillation length. In the standard
approach this is exactly the condition that the localization term
does not suppress the oscillations.

Finally we would like to note that the advantages of the discussed
approach are  physical clearness   and   technical simplicity.

\section*{Acknowledgments}
The authors are grateful to E. Boos, A. Lobanov, A. Pukhov, L.
Slad and Yu. Tchuvilsky for interesting and  useful discussions.
Special thanks are due to M. Smolyakov for reading the  manuscript
and making important comments. Analytical calculations of the
amplitudes have been carried out with the help of the COMPHEP and
REDUCE packages. The work of V. Egorov was supported by the
Foundation for the Advancement of Theoretical Physics and
Mathematics ``BASIS''.

\end{document}